\documentstyle[12pt,psfig]{article}
\makeatletter
\def\@cite#1#2{{[{#1}]\if@tempswa\typeout
{IJCGA warning: optional citation argument
ignored: `#2'} \fi}}

\newcount\@tempcntc
\def\@citex[#1]#2{\if@filesw\immediate\write\@auxout{\string\citation{#2}}\fi
 \@tempcnta\z@\@tempcntb\m@ne\def\@citea{}\@cite{\@for\@citeb:=#2\do
   {\@ifundefined
    {b@\@citeb}{\@citeo\@tempcntb\m@ne\@citea\def\@citea{,}{\bf ?}\@warning
     {Citation `\@citeb' on page \thepage \space undefined}}%
    {\setbox\z@\hbox{\global\@tempcntc0\csname b@\@citeb\endcsname\relax}%
    \ifnum\@tempcntc=\z@ \@citeo\@tempcntb\m@ne
   \@citea\def\@citea{,}\hbox{\csname b@\@citeb\endcsname}%
     \else
      \advance\@tempcntb\@ne
      \ifnum\@tempcntb=\@tempcntc
      \else\advance\@tempcntb\m@ne\@citeo
      \@tempcnta\@tempcntc\@tempcntb\@tempcntc\fi\fi}}\@citeo}{#1}}
\def\@citeo{\ifnum\@tempcnta>\@tempcntb\else\@citea\def\@citea{,}%
  \ifnum\@tempcnta=\@tempcntb\the\@tempcnta\else
   {\advance\@tempcnta\@ne\ifnum\@tempcnta=\@tempcntb \else
\def\@citea{--}\fi
   \advance\@tempcnta\m@ne\the\@tempcnta\@citea\the\@tempcntb}\fi\fi}
\makeatother

\def\boxit#1{\leavevmode\thinspace\hbox{\vrule\vtop{\vbox{\hrule%
        \vskip3pt\kern1pt\hbox{\vphantom{\bf/}\thinspace\thinspace%
        {\bf#1}\thinspace\thinspace}}\kern1pt\vskip3pt\hrule}\vrule}%
        \thinspace}
\def\Boxit#1{\noindent\vbox{\hrule\hbox{\vrule\kern3pt\vbox{
        \advance\hsize-7pt\vskip-\parskip\kern3pt\bf#1
        \hbox{\vrule height0pt depth\dp\strutbox width0pt}
        \kern3pt}\kern3pt\vrule}\hrule}}

\newcommand{\Hh}{\lower1.2ex\hbox{$\stackrel{\textstyle
H}{\footnotesize\sim}$}}
\newcommand{\Hho}{\lower1.2ex\hbox{$\stackrel{\textstyle
H_1}{\footnotesize\sim}$}}
\newcommand{\Hhw}{\lower1.2ex\hbox{$\stackrel{\textstyle
H_2}{\footnotesize\sim}$}}
\newcommand{\h}{\lower1.2ex\hbox{$\stackrel{\textstyle
h}{\footnotesize\sim}$}}

\newcommand{\gsim}{\lower.7ex\hbox{$\;\stackrel{\textstyle>}{\sim}\;$}}
\newcommand{\lsim}{\lower.7ex\hbox{$\;\stackrel{\textstyle<}{\sim}\;$}}
\newcommand{\be}{\begin{equation}}
\newcommand{\ee}{\end{equation}}
\newcommand{\bea}{\begin{eqnarray}}
\newcommand{\eea}{\end{eqnarray}}
\newcommand{\SUSY}{\makebox[1.25cm][l]{$\line(4,1){35}$\hspace{-1.15cm}{SUSY}}}
\newcommand{\tinySUSY}{\makebox[0.85cm][l]{$\line(4,1){19}$\hspace{-0.77cm}
{\tiny{SUSY}}}}

\def\simlt{\stackrel{<}{{}_\sim}}
\def\simgt{\stackrel{>}{{}_\sim}}

\def\baselinestretch{1}
\begin{document}
\catcode`@=11
\newtoks\@stequation
\def\subequations{\refstepcounter{equation}%
\edef\@savedequation{\the\c@equation}%
  \@stequation=\expandafter{\theequation}
  \edef\@savedtheequation{\the\@stequation}
  \edef\oldtheequation{\theequation}%
  \setcounter{equation}{0}%
  \def\theequation{\oldtheequation\alph{equation}}}
\def\endsubequations{\setcounter{equation}{\@savedequation}%
  \@stequation=\expandafter{\@savedtheequation}%
  \edef\theequation{\the\@stequation}\global\@ignoretrue

\noindent}
\catcode`@=12
\begin{titlepage}

\title{{\bf The MSSM fine tuning problem: a way out}}
\vskip3in
\author{  
{\bf J.A. Casas$^{1,2}$\footnote{\baselineskip=16pt E-mail address: {\tt
alberto@makoki.iem.csic.es}}},
{\bf J.R. Espinosa$^{2,3}$\footnote{\baselineskip=16pt E-mail address: {\tt
espinosa@makoki.iem.csic.es}}} and 
{\bf I. Hidalgo$^{2,3}$\footnote{\baselineskip=16pt E-mail address: {\tt
irene@makoki.iem.csic.es}}}
\hspace{3cm}\\
 $^{1}$~{\small I.E.M. (CSIC), Serrano 123, 28006 Madrid, Spain}.
\hspace{0.3cm}\\
 $^{2}$~{\small I.F.T. C-XVI, U.A.M., 28049 Madrid, Spain}.
\hspace{0.3cm}\\
 $^{3}$~{\small I.M.A.F.F. (CSIC), Serrano 113 bis, 28006 Madrid, Spain}.
} 
\date{} 
\maketitle 
\def\baselinestretch{1.15} 
\begin{abstract}
\noindent 
As is well known, electroweak breaking in the MSSM requires substantial
fine-tuning, mainly due to the smallness of the tree-level Higgs quartic
coupling, $\lambda_{\rm tree}$. Hence the fine tuning is efficiently
reduced in supersymmetric models with larger $\lambda_{\rm tree}$, as 
happens  naturally when the breaking of SUSY occurs at a low scale (not
far from the TeV). We show, in general and with specific examples, that a
dramatic improvement of the fine tuning (so that there is virtually no
fine-tuning) is indeed a very common feature of these scenarios for wide
ranges of $\tan \beta$ and the Higgs mass (which can be as large as
several hundred GeV if desired, but this is not necessary).  The
supersymmetric flavour problems are also drastically improved due to the
absence of RG cross-talk between soft mass parameters.
\end{abstract}

\thispagestyle{empty}
\vspace*{4cm}
\leftline{October 2003}
\leftline{}

\vskip-21cm
\rightline{IFT-UAM/CSIC-03-40}
\rightline{IEM-FT-233/03}
\rightline{hep-ph/0310137}
\vskip3in

\end{titlepage}
\setcounter{footnote}{0} \setcounter{page}{1}
\newpage
\baselineskip=20pt

\noindent

\section{The supersymmetric fine tuning problem}

One of the most attractive features of supersymmetry (SUSY) \cite{SUSY} is
that it provides a radiative mechanism for the electroweak (EW) breaking
\cite{Ibanez}. Large radiative corrections associated to the top Yukawa
coupling destabilize the origin of the Higgs potential and induce quite
naturally a non trivial minimum at the right scale, ${v}=246$ GeV, if the
mass terms that encode the soft breaking of SUSY are not far from the EW
scale.  This crucial success of SUSY has been undermined in recent times
by a worrisome fine tuning problem \cite{BG}-\cite{rest}: the non
observation of the Higgs boson and of superpartners sets significant lower
bounds on the size of the soft breaking terms in such a way that a
delicate cancellation is generically required to avoid too large a value
for $v$.

Let us briefly recall how this comes about in the ordinary Minimal
Supersymmetric Standard Model (MSSM). In the MSSM the Higgs 
sector
consists of two $SU(2)_L$ doublets, $H_1$, $H_2$. The 
(tree-level) scalar potential for the neutral components, $H^0_{1,2}$, 
of these doublets reads
\be
\label{VMSSM}
V^{\rm MSSM}(H^0_1,H^0_2)=m_1^2|H_1^0|^2+m_2^2|H_2^0|^2-2m_3^2
H_1^0 H^0_2 + \frac{1}{8}(g^2+g_Y^2) (|H^0_1|^2-|H^0_2|^2)^2,
\ee
with $m_{1,2}^2=\mu^2 + m_{H_{1,2}}^2$ and $m_3^2=B\mu$, where 
$m_{H_{i}}^2$ and $B$ are soft masses and $\mu$ is the
Higgs mass term in the superpotential, $W\supset \mu H_1\cdot  H_2$.
Minimization of $V^{\rm MSSM}$ leads to a vacuum expectation value (VEV)
$v^2\equiv 2(\langle H^0_1 \rangle^2 + \langle H^0_2\rangle^2)$
and thus to a mass for the $Z^0$ gauge boson, 
$M_Z^2=\frac{1}{4}(g^2+g_Y^2) v^2$, given by
\be
\label{MZMSSM}
\frac{M_Z^2}{2}=-\mu^2 + \frac{m_{H_{1}}^2-
m_{H_{2}}^2\tan^2\beta}{\tan^2\beta-1} \ ,
\ee
where $\tan\beta\equiv \langle H^0_2 \rangle/\langle H^0_1 \rangle$. The 
quantities in
the r.h.s. of (\ref{MZMSSM}) are to be understood as evaluated at low 
energy. They are related to more fundamental parameters at a higher scale,
$\Lambda_{UV}$ (typically $\Lambda_{UV}\equiv$ $M_{GUT}$ or $M_P$, but
there are other possibilities) by renormalization group equations
(RGEs). The MSSM RGEs for the mass parameters in (\ref{MZMSSM}) are
coupled to those of other soft terms, e.g. gaugino masses, stop masses,
trilinear terms, etc., so $M_Z^2$ can be expressed as a linear combination
of initial (UV) mass--squared parameters with coefficients that can be
calculated by integrating the RGEs. For example, for large $\tan\beta$
(the best situation for the fine tuning problem, as will be clear
in the discussion) and $\Lambda_{UV}=M_{GUT}=1.4\times 10^{16}$ GeV we get 
\cite{Howie}:
\be
\label{MZMSSM2}
M_Z^2\simeq -2.02\mu^2 + 3.57 M^2 + 0.07 m^2 + 0.22 A^2 + 0.75 A M\ ,
\ee
where $M,m,A$ are the gaugino mass, scalar soft mass and trilinear soft 
term respectively, taken universal for simplicity.  
From the previous equation it is 
apparent that, even 
for moderate values of the initial parameters (i.e. significantly smaller
than 1 TeV), some of the terms in the r.h.s. are much larger than
$M_Z^2$, thus a non-trivial cancellation among them (and therefore a
fine tuning) is necessary in general.

The previous fine tuning can be avoided in two ways. The first is that the
required cancellation among different terms is ``miraculously'' provided
by the fundamental theory underlying the MSSM, e.g. string theory. This
certainly would be a fortunate accident since the cancellation not only
involves the sizes of the various soft breaking terms (and the
$\mu$-parameter), which arise from the unknown SUSY breaking (\SUSY)
mechanism, but also the different magnitudes of the coefficients in
(\ref{MZMSSM2}), which have to do with the RG running between the initial
and the low energy scale. These quantities have such a different physical
origin that it is difficult to imagine a fundamental reason why they
should be correlated in the correct way to enforce a cancellation. As a
matter of fact, the analyses in the literature \cite{CEOP,mar,Gordy} of
many superstring, superstring-inspired and supergravity models do not find
such correlations.

The second way to avoid the fine tuning would be that each term in the 
r.h.s.~of 
(\ref{MZMSSM2}) is not larger than a few times $M_Z^2$. But, if the
soft masses are lowered at will, the masses of SUSY particles will fall
below their experimental bounds. The problem is especially acute for
the LEP bound on the Higgs mass, $m_h\ge 115$ GeV \cite{LEPbound} as has 
been stressed by
a number of authors \cite{LEPprob,LEPprob2,CEOP,Kane}. This can be easily 
understood 
by writing the tree-level and the dominant 1-loop correction \cite{mhrad} 
to the theoretical upper bound on $m_h$ in the MSSM:
\bea
\label{mhMSSM}
m_h^2\leq M_Z^2 \cos^2 2 \beta + {3 m_t^4 \over 2\pi^2 v^2}
\log{M_{\rm SUSY}^2\over m_t^2} + ...  \eea
where $m_t$ is the (running) top mass ($\simeq 167$ GeV for $M_t=174$ GeV)  
and $M_{\rm SUSY}$ is an average of stop masses. Since the experimental
lower bound on $m_h$ exceeds the tree-level contribution, the radiative
corrections must be responsible for the difference, and this translates
into a lower bound on $M_{\rm SUSY}$:
\bea
\label{MSUSYMSSM}
M_{\rm SUSY}\;\simgt\;  e^{-2.1\cos^2 2\beta}
 e^{\left({m_h}/{62\ {\rm GeV}}\right)^2} m_t\;\simgt\;   3.8\ m_t\ ,
\eea
where the last figure corresponds to $m_h= 115$ GeV and large $\tan
\beta$.  Hence $M_{\rm SUSY}^2$ must be already more than 40 times bigger
than $M_Z^2$ and this number increases exponentially for larger (smaller)  
$m_h$ ($\tan \beta$). On the other hand $M_{\rm SUSY}$ is itself a low
energy quantity that has a dependence on the initial soft masses 
analogous\footnote{We have approximated in Eq.~(\ref{MSMSSM2}) the
geometric average of the stop masses by the arithmetic one, which
is sufficiently precise for the argument.}
to eq.~(\ref{MZMSSM2}):
\be
\label{MSMSSM2}
M_{\rm SUSY}^2\simeq  3.36 M^2 + 0.49 m^2 - 0.05 A^2 - 0.19 A M + m_t^2 + 
({\rm D-terms})\ .
\ee
Roughly, $M_{\rm SUSY}^2$ has
a magnitude similar to the main positive contribution in the r.h.s. of
(\ref{MZMSSM2}), which then implies that some of the terms in that sum are
at least $\sim$ 45 times larger than $M_Z^2$, showing up the fine tuning. 
(The evaluation of the MSSM fine tuning is refined in the next section.) 

Several ways to alleviate the SUSY fine tuning problem have been explored
in the literature, {\em e.g.} invoking a
correlation between parameters based on some theoretical construction 
as mentioned above \cite{CEOP,mar,Gordy}. The
improvement, however, is never dramatic. Here we take a different path:  
the fine tuning can be alleviated or, indeed, eliminated if the
supersymmetric theory has larger tree-level quartic Higgs couplings than
the conventional one [see eq.~(\ref{VMSSM})]. In this way, the size of the
various contributions to $M_Z^2$ [eq.~(\ref{MZMSSM2}) for the MSSM] is
dramatically lowered and, besides, the soft breaking terms do not need to
be large since the radiative corrections are no longer necessary to
explain the Higgs mass. We show here (building on a previous observation 
in \cite{BCEN}) how this happens naturally if the SUSY
breaking occurs at a low scale (not far from the TeV).

The paper is organized as follows: In sect.~2 we discuss and identify the
causes for the abnormally large fine tuning of the MSSM, envisaging
possible cures. In sect.~3 we consider low scale \SUSY scenarios,
evaluating the corresponding fine tuning and showing that it can be
drastically smaller than in the MSSM. In sect.~4 we offer a concrete and
realistic realization of the mechanism in a specific model. In sect.~5 we
make some concluding remarks. Finally, in appendix~A we present formulas
for measuring the fine tuning in generic scenarios, while appendix~B
discusses the limits that Higgs searches at LEP impose on the parameters
of the kind of models we consider.

\section{Underlying causes and possible cures}

It is interesting to note that the fine tuning in the MSSM is much more
severe than what simple dimensional arguments suggest.  To show 
this we quantify the
fine tuning following Barbieri and Giudice \cite{BG}:  we write the Higgs
VEV as a function of the initial parameters  $p_\alpha$ of the model under 
study, $v^2=v^2(p_1, p_2, \cdots)$, and define $\Delta_{p_\alpha}$, the 
fine tuning
parameter associated to $p_\alpha$, by
\be
\label{BG}
{\delta M_Z^2\over M_Z^2}= {\delta v^2\over v^2} = 
\Delta_{p_\alpha}{\delta
p_\alpha\over p_\alpha}\ ,  
\ee
where $\delta M_Z^2$ ($\delta v^2$) is the change induced in $M_Z^2$
($v^2$) by a change $\delta p_\alpha$ in $p_\alpha$. Roughly speaking
$\Delta^{-1}_{p_\alpha}$ measures the probability of a cancellation among 
terms 
of a given size to obtain a result which is $\Delta_{p_\alpha}$ times 
smaller.\footnote{Strictly speaking, $\Delta_{p_\alpha}$ measures 
the sensitivity of $v^2$ against variations of $p_\alpha$, rather
than the degree of fine tuning \cite{dCC,Anderson}. However, for the EW 
breaking it is a perfectly reasonable fine tuning indicator 
\cite{dCC,Paolo}:
when $p_\alpha$ is a mass parameter, $\Delta_{p_\alpha}$ 
is large only around a cancellation point.} 
Absence of fine tuning requires that $\Delta_{p_\alpha}$ should not be 
larger that ${\cal O}(10)$.

The parameter that usually requires the largest fine tuning is $\mu^2$
because, due to the negative sign of its contribution in
eqs.~(\ref{MZMSSM}, \ref{MZMSSM2}), it has to compensate the (globally
positive and large) remaining contributions.\footnote{As pointed out in
ref.~\cite{Paolo}, it is more sensible to use $\mu^2$ rather than $\mu$ as
the parameter entering eq.~(\ref{BG}), since this is the form in which it
appears in the sum.} Therefore we will focus on $\Delta_{\mu^2}$.
For large $\tan\beta$ and $m_h=115$ GeV the required universal soft
mass is $m=M=A\simeq 325$ GeV, according to eqs.~(\ref{mhMSSM},
\ref{MSMSSM2}).  When these figures are plugged in eq.~(\ref{MZMSSM2})
and the fine tuning is evaluated according to eq.~(\ref{BG}) one
obtains $\Delta_{\mu^2}\sim 55$.  Note that this corresponds to using
the tree-level potential (\ref{VMSSM}), evaluated at a low scale.  If
one refines (\ref{VMSSM}) by including the dominant logarithmic
corrections at 1-loop from the top-stop sector (see appendix~A) this
figure gets down \cite{dCC,Poko} to $\sim 35$.\footnote{This estimate can be
softened further when subleading (1-loop and 2-loop) radiative
corrections  are added. The most important effect is related to the
one-loop  corrections from stop mixing. In the optimal case the figure
35 can be brought down to 20  \cite{LEPprob2,CEOP,Gordy2}.}

Now, one could naively expect that if the soft parameters had a size
$m_{\rm soft}^2\sim a v^2$, the fine tuning would be $\Delta\sim a$, but
for the MSSM one gets $\Delta\simgt 20a$, as can be checked from the
previous numbers. In this sense the fine tuning of the MSSM is
abnormally large. To understand the reasons for this, let us write the
generic Higgs potential along the breaking direction as
\be
\label{Vbeta}
V={1 \over 2} m^2 v^2 + {1 \over 4} \lambda v^4 \ ,
\ee
where $\lambda$ and $m^2$ are functions of the $p_\alpha$ parameters and 
$\tan \beta$, in particular
\be
\label{cimi}
m^2 = 
c_\beta^2\, m_1^2(p_\alpha)+s_\beta^2\, m_2^2(p_\alpha)-s_{2\beta}\, 
m_3^2(p_\alpha)\ .
\ee
Minimization of (\ref{Vbeta}) leads to
\be
\label{v2}
v^2={-m^2\over \lambda} \ .
\ee
Clearly, the larger the size of the individual $m_i^2$ and 
the smaller $\lambda$, the more
severe the fine tuning: $\Delta \sim {\tt m_i^2}/(\lambda v^2)$,
where ${\tt m_i^2}$ are the (potentially large) individual
contributions to $m_i^2$.  For
the MSSM, $\lambda$ turns out to be quite small:
\be
\label{lambdaMSSM}
\lambda_{\rm MSSM}={1 \over 8}(g^2+g_Y^2)\cos^2 2\beta\ \simeq \  {1
\over 15}\cos^2 2\beta\ ,
\ee
which already implies a fine tuning $\sim 15$ times larger (for the 
most favorable case of large $\tan \beta$) than the above
naive expectations.  
The previous $\lambda_{\rm MSSM}$ was evaluated at tree-level but radiative
corrections make $\lambda$ larger, thus reducing the fine tuning.  
The ratio $\lambda_{\rm tree}/\lambda_{\rm 1-loop}$ is basically
the ratio $(m_h^2)_{\rm tree}/(m_h^2)_{\rm 1-loop}$, so for large $\tan
\beta$ and $m_h=115$ GeV the previous factor 15 is reduced by a factor
$M_Z^2/m_h^2$ down to $\sim 9$.  Finally, for the MSSM (with large $\tan
\beta$ and $\Lambda_{UV}=M_{GUT}$)
\be
\label{mMSSM}
m^2=m_1^2 c^2_\beta+m_2^2 s^2_\beta - m_3^2 s_{2\beta}
\simeq 1.01\mu^2-2.31 \tilde{m}^2\ ,
\ee
where we have set $A=M=m=\tilde{m}$ for simplicity.
The presence of a sizeable RG coefficient in front of $\tilde{m}^2$ 
implies that, for a given magnitude of the latter, the 
required cancellation must be (in this case) $\sim 2.31$ times
more accurate than naive expectations so, finally the factor 9 is 
enhanced to $\sim 20$. 
Notice that those large RG coefficients are a consequence 
of the radiative mechanism of EW breaking, hence if the EW breaking 
were at tree level the fine tuning would be reduced.

From the above discussion it is important to notice that, although for a
given size of the soft terms the radiative corrections reduce the fine
tuning, the requirement of sizeable radiative corrections implies itself
large soft terms, which in turn worsens the fine tuning. More precisely,
for the MSSM $\delta_{\rm rad}\lambda\propto\log(M_{\rm SUSY}/m_t)$, so
$\lambda$ can only be radiatively enhanced by increasing $M_{\rm SUSY}$,
and thus the individual ${m}_i^2$. A given increase in $M_{\rm SUSY}$
reflects linearly in ${m}_i^2$ and only logarithmically in $\lambda$, so
the fine tuning $\Delta \sim {\tt m_i^2}/(\lambda v^2)$ gets usually worse. As
discussed in sect.~1, for the MSSM $(m_h)_{\rm tree}<(m_h)_{\rm exp}$,
hence sizeable radiative corrections are in fact mandatory and the fine
tuning is consequently aggravated. As a consequence, the fine tuning
increases exponentially for increasing (decreasing) $m_h$ ($\cos^22\beta$)
as indicated by eq.~(\ref{MSUSYMSSM}).

\begin{figure}[t]
\vspace{1.cm} \centerline{
\psfig{figure=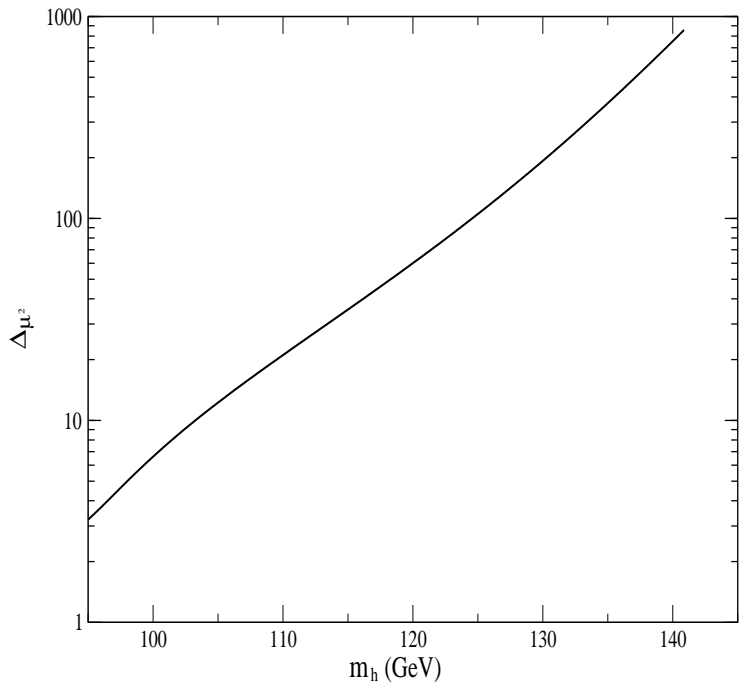,height=8cm,width=8cm,angle=-90,bbllx=5.cm,%
bblly=4.cm,bburx=20.cm,bbury=24.cm}}
\caption{\footnotesize 
Fine tuning in the MSSM (measured by $\Delta_{\mu^2}$) as a function of 
the Higgs mass (in GeV) for $\tan\beta=10$.}
\label{ftMSSMvsmh}
\end{figure}

\begin{figure}[t]
\vspace{1.cm} \centerline{
\psfig{figure=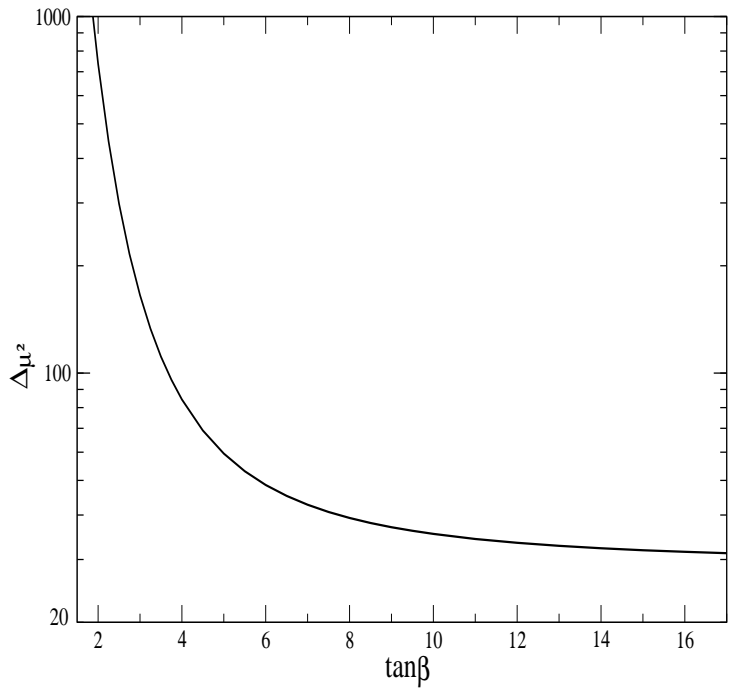,height=8cm,width=8cm,angle=-90,bbllx=5.cm,%
bblly=4.cm,bburx=20.cm,bbury=24.cm}}
\caption{\footnotesize
Lower bound on the MSSM fine tuning ($\Delta_{\mu^2}$) as a function of
$\tan\beta$ from the LEP bound $m_h\geq 115$ GeV.}
\label{LEPprice}
\end{figure}
                                                                                
\begin{figure}[t]
\vspace{1.cm} \centerline{
\psfig{figure=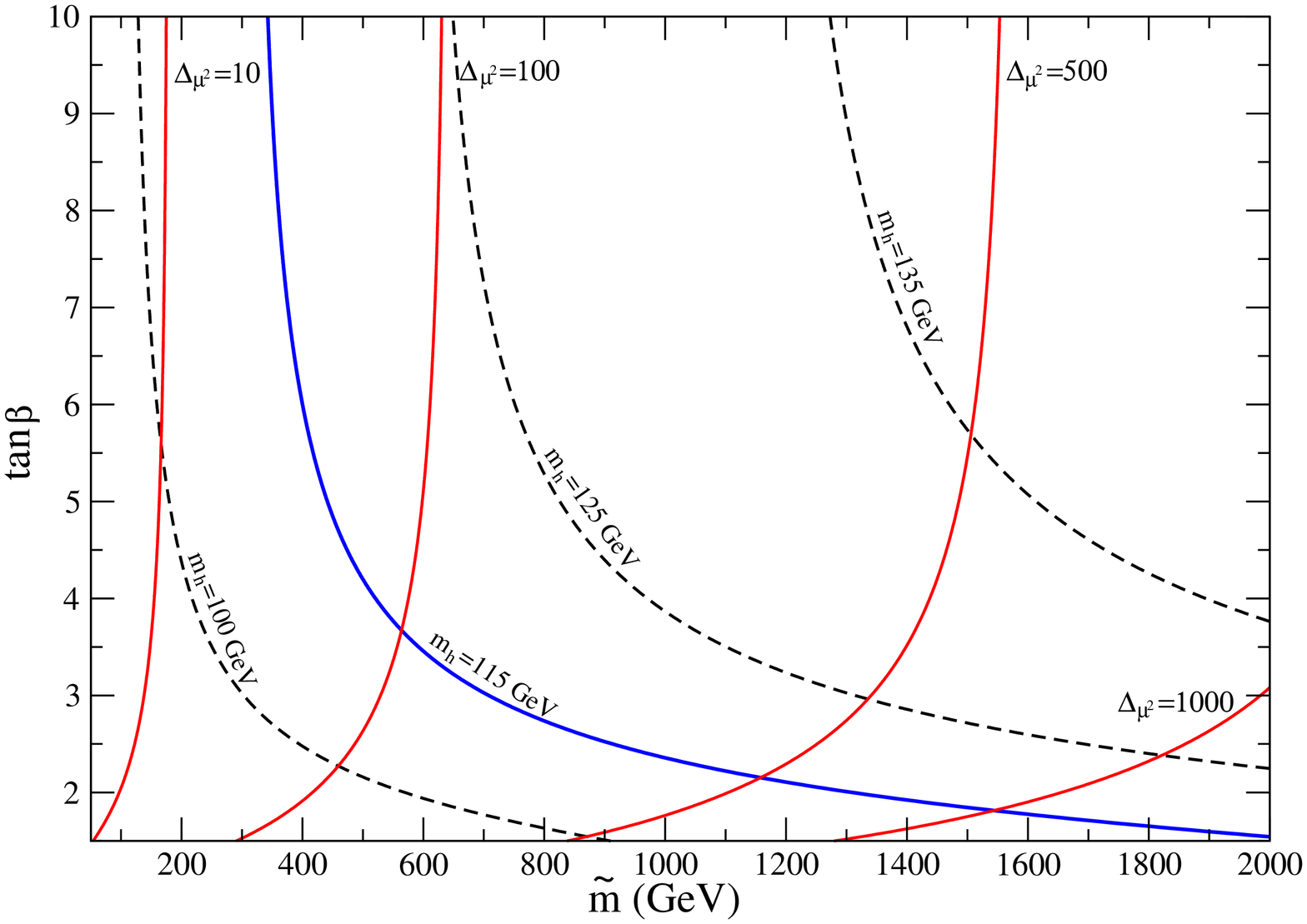,height=8cm,width=8cm,angle=-90,bbllx=5.cm,%
bblly=4.cm,bburx=20.cm,bbury=24.cm}}
\caption{\footnotesize
Fine tuning in the MSSM (measured by $\Delta_{\mu^2}$, solid lines) in the
$(\tilde{m},\tan\beta)$ plane. Dashed lines are contour lines of constant
Higgs mass.}
\label{ftMSSM}
\end{figure}

Let us illustrate the previous discussion in a more quantitative way.
In fig.~\ref{ftMSSMvsmh} we plot $\Delta_{\mu^2}$, evaluated at one-loop,
as a function of the Higgs boson mass, $m_h$, for $\tan\beta=10$ (such
large value of $\tan\beta$ minimizes the fine tuning, as discussed above).
We only include the dominant one-loop correction to $m_h$, as shown in
eq.~(\ref{mhMSSM}), and make the simplifying assumption that the soft
parameters are universal at the GUT scale. Although the fine tuning can be
made smaller in non-universal cases, figure~\ref{ftMSSMvsmh} shows the 
typical size of $\Delta_{\mu^2}$ in the MSSM. As expected from the previous 
discussions, $\Delta_{\mu^2}$ grows exponentially for increasing $m_h$. The 
dependence
of $\Delta_{\mu^2}$ with $\tan\beta$ is shown in fig.~\ref{LEPprice} for
$m_h$ at the LEP bound\footnote{With
our choice of universal soft masses, the mass of the pseudoscalar Higgs is
generically large. In that case the LEP bound reduces to that in the SM:
$m_h\simgt 115$ GeV.}, $m_h=115$ GeV (the optimal choice for the fine
tuning). The curve  for $\Delta_{\mu^2}$ increases exponentially for
decreasing $\cos^2 2\beta$, again as expected. This 
curve can be interpreted as a LEP lower bound on the MSSM fine tuning.

Finally, fig.~\ref{ftMSSM} shows contour lines of constant
$\Delta_{\mu^2}$ in the $(\tilde{m},\tan\beta)$ plane, where $\tilde{m}$
is the universal soft mass at $\Lambda_{UV}$. We also plot dashed contour
lines of constant $m_h$ and the LEP lower bound on $m_h$. Again, it is 
clear how the fine tuning is greater
for smaller $\tan\beta$ and how it grows, together with $m_h$, for larger
$\tilde{m}$. The upper horizontal line and the $m_h=115$ GeV contour line
correspond to figs.~\ref{ftMSSMvsmh} and \ref{LEPprice} respectively.  It
is instructive to examine the behaviour of the lines of constant
$\Delta_{\mu^2}$ along which $\tilde{m}$ and $m_h$ grow asymptotically
towards fixed upper limits. For instance, if we insist in having small
fine tuning, $\Delta_{\mu^2}\simlt 10$, following the $\Delta_{\mu^2}=10$
line we conclude that one cannot obtain $\tilde{m}$ larger than
$\sim 175$ GeV (which translates into upper bounds on superpartner masses)  
nor Higgs masses larger than $\simeq 103$ GeV, already ruled out by LEP.

A word of caution should be added about the previous numbers. In general,
attempting a very accurate determination of the fine tuning does not make
much sense. What precise value of the fine tuning should be considered
too high? On top of this, the present experimental uncertainty on the top
quark mass, $M_t=174.3\pm 5.1$ GeV \cite{top}, translates into a 
significant
uncertainty on the fine tuning parameters.\footnote{{\em E.g.},  fixing 
$\tan\beta=10$ and $m_h=115$
GeV, one gets $\Delta_{\mu^2}=35^{-7}_{+12}$ for $M_t=174.3\pm 5.1$ 
GeV.} For these
reasons, in our numerical one-loop estimates of $\Delta_{\mu^2}$ we have
just included the logarithmic correction to $m_h^2$ given in
eq.~(\ref{mhMSSM}).  This simplification is even more justified in this 
paper, whose
main purpose is to compare the performance of unconventional scenarios 
with that of the MSSM.

In summary, the fine tuning of the MSSM is at least 20 times more severe
than naively expected due, basically, to the smallness of the tree-level
Higgs quartic coupling, $\lambda_{\rm tree}$. The problem is worsened by 
the fact that sizeable
radiative corrections (and thus sizeable soft terms) are needed to satisfy
the experimental bound on $m_h$. This is also due the smallness of
$\lambda_{\rm tree}$: if it were bigger, radiative corrections would not be
necessary.  In consequence, the most efficient way of reducing the fine
tuning is to consider supersymmetric models where $\lambda_{\rm tree}$ is
larger than in the MSSM. More explicitly, the improvement can be evaluated
in the following way. The value of $\Delta_p$ for a generic  parameter $p$
of a given model has the form [see Appendix~A]
\bea
\label{Deltap}
\Delta_p = \frac{p}{m^2}\left[\frac{\partial
m^2}{\partial p} + \frac{v^2}{2}\frac{\partial \lambda}{\partial
\beta}\frac{d \beta} {d p} + v^2\frac{\partial
\lambda}{\partial p} \right] \ .
\eea
Focusing on the $\mu^2$ parameter, and taking into account that the
last two terms of (\ref{Deltap}) are usually suppressed by a factor ${\cal 
O}(v^2/\mu^2)$, we can write
\bea
\label{Deltamu2}
\Delta_{\mu^2}\simeq \frac{\mu^2}{m^2}\frac{\partial
m^2}{\partial \mu^2}\simeq -\frac{\mu^2}{\lambda v^2} \simeq
-2\frac{\mu^2}{m_h^2} \ ,
\eea
where we have used the fact that the dependence of the low-energy
$m^2$ on the initial (UV) $\mu$ parameter is usually
dominated by the tree-level contribution. Strictly speaking, $m_h^2$ in 
(\ref{Deltamu2}) is the Higgs mass matrix element along the breaking 
direction, but in many cases of interest it is very close to one of the 
mass eigenvalues.
Therefore
\bea
\label{Deltamu2s}
\Delta_{\mu^2}\simeq \Delta_{\mu^2}^{\rm MSSM}
\left[\frac{m_h^{\rm MSSM}}{m_h} \right]^2 
\left[\frac{\mu}{\mu^{\rm MSSM}} \right]^2 \ .
\eea
This equation shows the two main ways in which a theory can improve
the MSSM fine tuning: increasing $m_h$ and/or decreasing $\mu$. The
first way corresponds to increasing $\lambda$. The second, for
a given $m_h$, corresponds to reducing the size of the soft
terms (EW breaking requires the size of $\mu^2$ to be
proportional to the overall size of the soft squared-masses), which is 
only allowed if radiative contributions are not essential to raise
$m_h$. Both improvements indeed concur for larger $\lambda_{\rm 
tree}$.

The possibility of having tree-level quartic Higgs couplings larger than
in the MSSM is natural in scenarios in which the breaking of SUSY occurs
at a low-scale (not far from the TeV scale)  
\cite{hard,Brignole,Polonsky,BCEN}.\footnote{This can also happen in 
models with extra dimensions opening up not far from the electroweak scale 
\cite{Strumia}. Another way of increasing
$\lambda_{\rm tree}$ is to extend the gauge sector \cite{extG} or to
enlarge the Higgs sector \cite{extH}. The latter option has been studied
in \cite{NMSSM} (for the NMSSM) but this framework is less effective in
our opinion, see sect.~5.} Besides, in that framework EW breaking takes
place essentially at tree-level, which, as noticed above, is also welcome
for the fine tuning issue. These ideas are developed in detail in the next
sections.

\section{Low-scale SUSY breaking}

In any realistic breaking of SUSY, there are two scales involved: 
the \SUSY scale, say $\sqrt{F}$, which corresponds
to the VEVs of the relevant auxiliary fields in the \SUSY sector; and the
messenger scale, $M$, associated to the interactions that transmit the
breaking (through effective operators suppressed by powers of $M$) to the 
observable sector.
These operators give rise to soft terms (such as scalar soft masses), but
also hard terms (such as quartic scalar couplings):
\be
\label{mlambda}
m_{\rm soft}^2\sim {F^2\over M^2}\ ,\;\;\;\; \lambda_{\tinySUSY}\sim
{F^2\over M^4}\sim {m_{\rm soft}^2\over M^2}\ .
\ee
Phenomenology requires $m_{\rm soft} = {\cal O}(1\, {\rm TeV})$, but 
this does not
fix the scales $\sqrt{F}$ and $M$ separately.  The MSSM assumption is that 
there is 
a
hierarchy of scales: $m_{\rm soft}\ll \sqrt{F}\ll M$, so that the
hard terms are negligible and the soft ones are the only observable trace
of \SUSY. However, there is no real need for such a strong hierarchy, so
the scales $\sqrt{F}$ and $M$  could well be  of 
similar order (thus not far from the TeV scale).  This happens in the 
so-called
low-scale \SUSY scenarios.  In this framework, the hard terms of
eq.~(\ref{mlambda}), are not negligible anymore and hence the \SUSY
contributions to the Higgs quartic couplings can be easily larger than the
ordinary MSSM value (\ref{lambdaMSSM}).  As discussed in the previous 
section, this 
is
exactly the optimal situation to ameliorate the fine tuning problem.

The messenger scale $M$ may be not far from the EW scale for
various reasons. E.g. there could be some massive fields responsible for
the \SUSY mediation (like in gauge mediation) with masses $\sim M$; or
there could be a more fundamental reason, as in models with large
extra-dimensions or in supersymmetric Randall-Sundrum models. 
Instead of sticking to one of these particular examples, it is convenient 
to describe the observable
physics using an effective field-theory approach \cite{Brignole,BCEN}. 
Denoting by
$T$ the superfield responsible for \SUSY, $\langle F_T\rangle\neq 0$, and
assuming that, apart from the $T$ field, the spectrum is minimal (i.e. the
same as in the MSSM), the effective theory is like the SUSY part of the 
MSSM, plus some effective interactions which include couplings between $T$ 
and the observable fields, suppressed by powers of $M$. These effective
interactions can appear in the superpotential, $W$, as well as in the
K\"ahler potential, $K$, or the gauge kinetic function.

\begin{figure}[t]
\vspace{1.cm} \centerline{
\psfig{figure=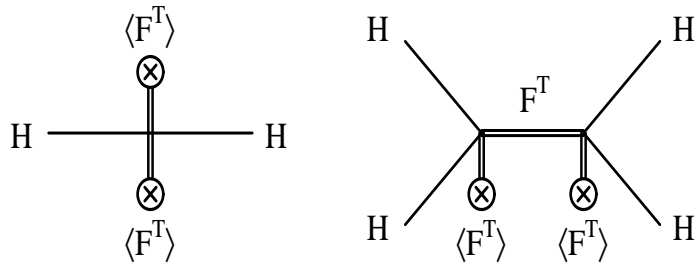,height=4cm,width=6cm,bbllx=8.cm,%
bblly=19.cm,bburx=17.cm,bbury=24.cm}}
\caption{\footnotesize  Higgs soft masses and hard quartic couplings
that arise from the K\"ahler operator (\ref{K1}).}
\label{diag}
\end{figure}

As a simple example, suppose that the K\"ahler potential contains the 
operator
\bea
\label{K1}
K\ \supset\  -\frac{1}{M^2}|T|^2|H|^2 + \cdots \eea
where $H$ denotes any
Higgs superfield. Once $F_T$ takes a VEV the above nonrenormalizable
interaction produces soft terms as well as hard terms, as 
schematically  represented in the diagrams of Fig.~\ref{diag}. 
Notice that $m_{\rm soft}^2\sim |F_T|^2/M^2$, 
$\lambda_{\tinySUSY}\sim |F_T|^2/M^4\sim 
m_{\rm soft}^2/M^2$, in agreement with (\ref{mlambda}).

In general, 
the Higgs potential has the structure 
of a generic two Higgs doublet model (2HDM), with  $T$-dependent 
coefficients
\cite{BCEN},
\bea
\label{VTH}
V & = &  V_0(\bar{T},T) + m_1^2(\bar{T},T)|H_1|^2  +
m_2^2(\bar{T},T)|H_2|^2  - \left[ m_3^2(\bar{T},T) H_1\cdot H_2 + {\rm
h.c.}\right] \nonumber \\ & + &   {1\over 2}
\lambda_1(\bar{T},T)|H_1|^4  +{1\over 2} \lambda_2(\bar{T},T)|H_2|^4
+\lambda_3(\bar{T},T)|H_1|^2 |H_2|^2  +\lambda_4(\bar{T},T)|H_1\cdot
H_2|^2  \nonumber \\ & + &  \left[ {1 \over 2} \lambda_5(\bar{T},T)
(H_1\cdot H_2)^2  +\lambda_6(\bar{T},T)|H_1|^2 H_1\cdot H_2
+\lambda_7(\bar{T},T)|H_2|^2 H_1\cdot H_2  +  {\rm h.c.}
\right]\nonumber\\ &+&\ldots 
\eea
where we have truncated at ${\cal O}(H^4)$, which makes sense whenever
$v^2/M^2$ is small. The quantities $m_i^2$, $\lambda_i$ can be expressed
in terms of the parameters appearing in $W$ and $K$ (explicit expressions 
can be found in ref.~\cite{BCEN}). If the $T$ field is heavy enough it can 
be 
integrated out and 
one ends up with a truly 2HDM. The previous potential is to be compared 
with the MSSM one [eq.~(\ref{VMSSM})] with $\lambda_{1,2}={1\over 
4}(g^2 + g_Y^2)$, $\lambda_3={1\over
4}(g^2 - g_Y^2)$,  $\lambda_4=-{1\over 2}g^2$, $\lambda_{5,6,7}=0$.

The appearance of non-conventional quartic couplings has a deep impact
on the pattern of EW breaking \cite{BCEN}. In the MSSM, the existence of 
D-flat
directions, $|H_1|=|H_2|$, imposes the well-known condition, $m_1^2 +
m_2^2 - 2 |m_3^2| > 0$, in order to avoid a potential unbounded from
below along such directions. However, the boundedness of the potential can
now be simply ensured by the contribution of the extra quartic
couplings, and this opens up many new possibilities for EW breaking.  For
example, the universal case $m_1^2= m_2^2$ is now allowed, as well as
the possibility of having both $m_1^2$ and $m_2^2$ negative (with
$m_3^2$ playing a minor role).  In addition, and unlike in the MSSM, there 
is no need of radiative corrections to destabilize the origin, and EW 
breaking generically occurs already at tree-level (which is just fine
 since the effects of the RG running are
small as the cut-off scale is $M$).
Moreover, this tree-level breaking (which is welcome for the fine
tuning issue, as discussed in sect.~2) occurs naturally only in the
Higgs sector \cite{BCEN}, as desired.

Finally, the fact that quartic couplings are very different from those
of the MSSM changes dramatically the Higgs spectrum and properties.
In particular, the MSSM upper bound on the mass of the lightest
Higgs field no
longer applies, which has also an important and positive impact on
the fine tuning problem, as is clear from the discussion after 
eq.~(\ref{Deltamu2s}).

\section{A concrete model}

In this section we evaluate numerically the fine tuning involved in the EW 
symmetry breaking in a particular model with low-scale \SUSY and compare 
it with that of the MSSM. We choose a model first introduced (as 
"example A") in 
\cite{BCEN}  and analyzed there for its own sake. We show 
now that the fine tuning problem is greatly softened in this model 
even if it was not constructed with that goal in mind.

The superpotential is given by
\be 
W =\Lambda_S^2 T + \mu H_1\cdot H_2 + {\ell\over 2M}(H_1\cdot H_2)^2 \ ,
\ee
and the K\"ahler potential is
\bea 
K & = & |T|^2 +  |H_1|^2 +  |H_2|^2\nonumber\\ 
& - & {\alpha_t \over 4 M^2} |T|^4 + {\alpha_1 \over
M^2}|T|^2  \left(|H_1|^2+|H_2|^2\right) + {e_1 \over
2M^2}\left(|H_1|^4+|H_2|^4\right)\ .
\eea
(All parameters are real with $\alpha_t>0$.) Here $T$ is the 
singlet field responsible for the breaking of supersymmetry, $\Lambda_S$ 
is the \SUSY scale and $M$ the `messenger' scale (see previous section).
The typical soft masses are $\sim \tilde{m}\equiv \Lambda_S^2/M$. In 
particular, the mass of the scalar component of $T$ is 
${\cal{O}}(\tilde{m})$ and, after integrating this field out, the 
effective potential for $H_1$ and $H_2$ is a 2HDM, like (\ref{VTH}), with
very particular Higgs mass terms:
\be 
\label{softm}
m_1^2=m_2^2=\mu^2-\alpha_1 \tilde{m}^2\ ,\;\;\;\; m_3^2=0\ , 
\ee
and Higgs quartic
couplings like those of the MSSM plus contributions of order $\mu/M$ and
$\tilde{m}^2/M^2$:
\bea
\label{quarticc}
\lambda_1=\lambda_2&=&{1\over 4}(g^2+g_Y^2)+2\alpha_1^2{\tilde{m}^2\over
M^2}\ ,\nonumber\\ \lambda_3&=&{1\over 4}(g^2-g_Y^2)+{2\over
M^2}(\alpha_1^2\tilde{m}^2-e_1\mu^2)\ ,\nonumber\\  \lambda_4&=&-{1\over
2}g^2-2\left(e_1+2{\alpha_1^2\over \alpha_t}\right){\mu^2\over M^2}\
,\nonumber\\  \lambda_5&=&0\ ,\nonumber\\
\lambda_6=\lambda_7&=&{\ell\mu\over M}\ .
\eea
The symmetry of the potential under $H_1\leftrightarrow H_2$ allows to 
solve the minimization conditions explicitly not only for $v$ 
but also for $\tan\beta$. Depending on the value of 
the parameter $l$, one gets either\footnote{One has  
${\rm sgn}(\tan\beta)=-{\rm sgn}(l\mu/M)$. We are 
implicitly taking parameters such that $\tan\beta>0$.} $\tan\beta=1$ or 
$\tan\beta>1$. The explicit 
expressions for $v$ and $\sin2\beta$, and the spectrum of Higgs masses, 
can be found in \cite{BCEN}. 
One important difference with respect to the 
MSSM spectrum is that all Higgs masses are now of order $v$. The CP-even 
scalars $h,H$ can be in the region accessible to LEP searches. Although 
the charged Higgs, $H^\pm$, and the pseudoscalar, $A^0$, are
usually too heavy for detection at LEP, in some regions of
parameter space they might also be light and their possible production 
must be considered too. Limits on the parameter space
of this model that result from Higgs searches at LEP are discussed in
Appendix~B and will be explicitly shown later on. 

To evaluate the fine tuning in this model we simply
plug (\ref{softm}) and (\ref{quarticc}) in the general expression for 
$\Delta_{\mu^2}$ given in Appendix~A [eq.~(\ref{Deltap2A})] to obtain
\be
\Delta_{\mu^2}=-{\mu^2\over \lambda v^2}\left[1+v^2\left(
{ls_{2\beta}\over 2\mu M}-{1\over M^2}\hat{e}_1s_{2\beta}^2
\right)\right]\ ,
\label{newfine}
\ee
where $\lambda$ is the quartic scalar coupling along the breaking 
direction, 
explicitly given in eq.~(\ref{dilambdaiA}) and  $\hat{e}_1\equiv 
e_1+\alpha_1^2/\alpha_t$. This expression is 
valid
both for $\tan\beta=1$ and $\tan\beta>1$ and as discussed in 
Sect.~2 is dominated by the first term. Although (\ref{newfine}) is a
tree-level result, useful for understanding most of the parametric
dependence of $\Delta_{\mu^2}$, we use for the numerical comparison with
the MSSM a one-loop-refined evaluation of $\Delta_{\mu^2}$ (both in the
MSSM and the present model), computed following the procedure explained in
Appendix~A. 

We should also comment on the relation between the coupling 
$\lambda$ along the breaking direction [which is the coupling relevant for 
(\ref{newfine})] and the Higgs mass.
At tree-level one of the CP-even Higgses lies 
along the breaking direction and therefore has 
mass-squared $2\lambda v^2$, but this is no longer the case at one 
loop: radiative corrections induce a deviation in the direction of 
the mass eigenstates, the effect being larger for light tree level 
masses.  
We will use the notation $m_\parallel^2=2\lambda v^2$ for the mass matrix 
element 
that controls the fine tuning (\ref{newfine}) keeping in mind that it does 
not always correspond 
to the mass of a physical state. Explicitly, in the region $\tan\beta>1$
on which we focus here,
\be
m_\parallel^2=\left[{1\over 4}(g^2+g_Y^2)+2\alpha_1^2{\tilde{m}^2\over 
M^2}+{l\mu\over M}s_{2\beta}\right]v^2\ +
 {3m_t^4 \over 2\pi^2 v^2}
\log{M_{\rm SUSY}^2\over m_t^2} + ...
\ee
where we have added the dominant one-loop stop correction, as in the MSSM.

For the values of the parameters of the unconventional model we take as
a first example (set A) those used in \cite{BCEN}: $\mu/M=0.6$,
$e_1=-1.3$, $\tilde{m}/M=0.5$ and $\alpha_t=3$. The exact value of
$\alpha_1$ is fixed by the minimization condition for $v$: it is always
$\alpha_1\simgt \mu^2/\tilde{m}^2=1.44$, and gets closer and closer to
$\mu^2/\tilde{m}^2$ for increasing $\tilde{m}$. The parameter $l$ is
free and can be traded by $\tan\beta$. 
To understand some of the numerical results
that follow it is important to study the dependence of $m_\parallel^2$
on $\tilde{m}$ (for fixed $\tilde{m}/M$). Its tree level part 
decreases monotonically with increasing 
$\tilde{m}$ due to the behaviour of $\alpha_1$, while the one-loop 
correction
increases logarithmically with $\tilde{m}$ (it enters through
$M^2_{\rm SUSY}$ which we take to be $M_{\rm 
SUSY}^2\simeq \tilde{m}^2+m_t^2$). 
The combination of these two opposite effects results
in a mass $m_\parallel$ that decreases with $\tilde{m}$ for small
$\tilde{m}$ (where the tree-level dependence dominates), reaches a 
minimum,
and then starts increasing again for larger $\tilde{m}$ (when the
one-loop dependence takes over). For this reason, every value of 
$m_\parallel$ corresponds  to two values of
$\tilde{m}$: a low value, associated to a large tree level Higgs
coupling and a small radiative effect, which has small fine tuning; and a 
high value, associated to a larger radiative effect, which has larger fine 
tuning. This causes $\Delta_{\mu^2}$ to be a bi-valued function of 
$m_\parallel$. Moreover, for this set of parameters 
$m_\parallel$ is a good approximation to $m_h$.

\begin{figure}[t]
\vspace{1.cm} \centerline{
\psfig{figure=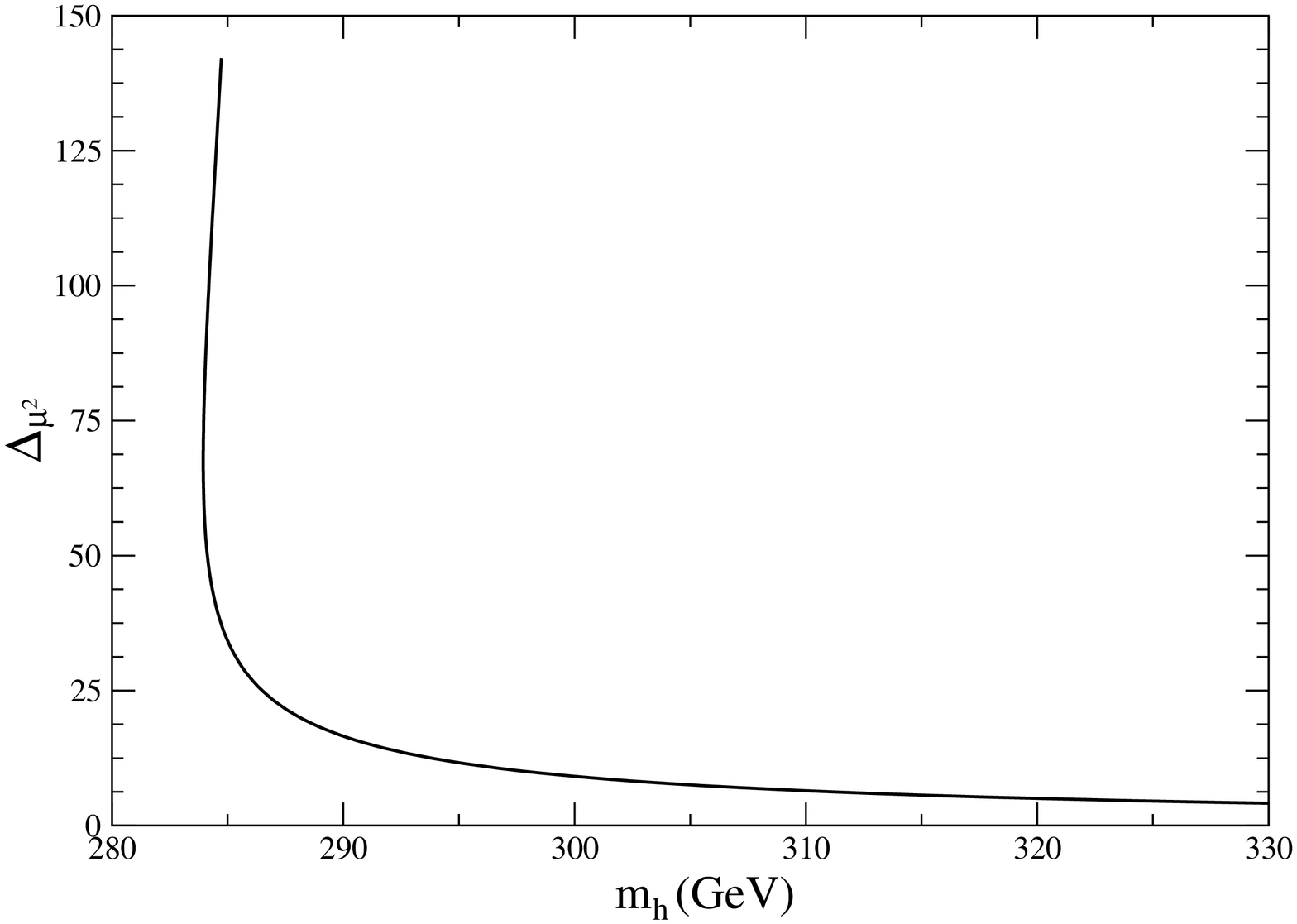,angle=-90,height=6cm,width=5cm,bbllx=5.cm,%
bblly=9.cm,bburx=20.cm,bbury=26.cm}
\psfig{figure=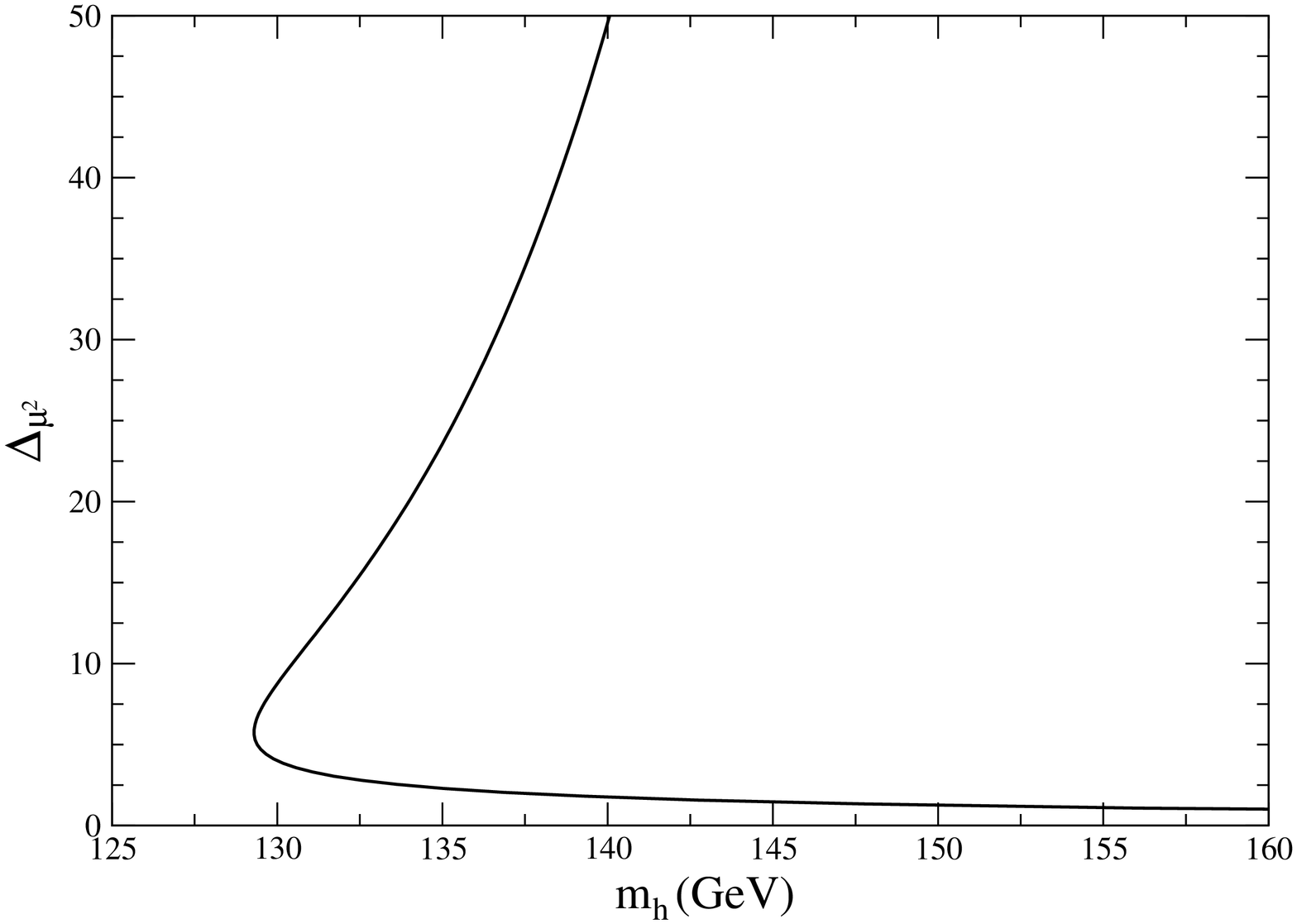,angle=-90,height=6cm,width=5cm,bbllx=5.cm,%
bblly=1.cm,bburx=20.cm,bbury=18.cm}
}
\caption{\footnotesize 
Fine tuning in the unconventional SUSY scenario of section~4 as a function
of the Higgs mass (in GeV) for $\tan\beta=10$ and the rest of parameters
as in set~A (left) or as in set~B (right).}
\label{ftnewvsmh}
\end{figure}

This behaviour is shown in Figure~\ref{ftnewvsmh}, left plot, which is the 
equivalent of fig.~\ref{ftMSSMvsmh}, but for the unconventional scenario 
just introduced, with $\tan\beta=10$. 
We can use the soft mass $\tilde{m}$ as a parameter along the curve
plotted, with $\Delta_{\mu^2}$ growing for increasing $\tilde{m}$.  In
the large-$\tilde{m}$ range of this curve (its steep upper branch)  
radiative corrections dominate the Higgs mass and the behaviour of the
fine tuning is similar to that in the MSSM (i.e. it grows with
increasing $m_h$). If we restrict our attention to the more interesting
low-$\tilde{m}$ range (the lower branch of the curve), the contrast with
the MSSM result is evident: now, the larger $m_h$ is, the smaller the
tuning becomes and for $m_h\simgt 300$ one gets $\Delta_{\mu^2}<10$. All 
this
is the straightforward result of having a larger tree level
contribution to the Higgs mass. For the choice of parameters considered
here (set A) the resulting Higgs mass is somewhat large, but we can
easily choose different parameters in order to lower the Higgs mass
without loosing the dramatic improvement in $\Delta_{\mu^2}$. This is
shown on the right plot of fig.~\ref{ftnewvsmh}, which has (set B):  
$\mu/M=0.3$, $e_1=-2$, $\tilde{m}/M=0.5$, $\alpha_t=1$ and
$\alpha_1\simgt 0.36$. The bi-valuedness of $\Delta_{\mu^2}$ is
more evident in this case.
\begin{figure}[t] 
\vspace{1.cm} 
\centerline{
\psfig{figure=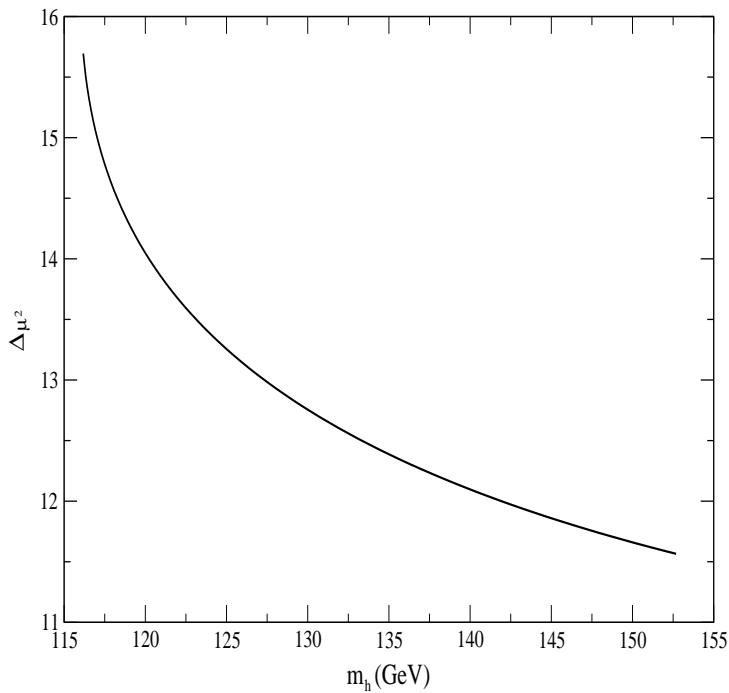,height=8cm,width=8cm,angle=-90,bbllx=5.cm,%
bblly=4.cm,bburx=20.cm,bbury=24.cm}} 
\caption{\footnotesize Fine tuning
in a low-scale SUSY breaking 
scenario as a function of the Higgs 
mass (in GeV) for $\tan\beta=10$.}
\label{ftnewvsmh2} 
\end{figure}

We plot  $\Delta_{\mu^2}$ vs. $m_h$ in Fig.~\ref{ftnewvsmh2} to make even 
clearer the difference of behaviour with respect to the MSSM (see 
fig.~\ref{ftMSSMvsmh}). We take $\mu= 330$ GeV, $\tilde{m}=550$ GeV, 
$e_1=-2$, $\alpha_t=1$, $l$ chosen 
to give $\tan\beta=10$ and instead of 
fixing $\tilde{m}/M$ we vary it from 0.05 to 0.8. In this way we can study 
the
effect on the fine tuning of varying $\lambda$ when the low energy mass
scales ($\mu$ and $\tilde{m}$) are kept fixed. When $\tilde{m}/M$ is small
(and this implies that $\mu/M$ is also small), the unconventional
corrections to quartic couplings are not very important and the Higgs mass
tends to its MSSM value\footnote{For the model at hand this limit is not
realistic, as it implies too small (or even negative) values of
$m_A^2$, $m_H^2$ and $m_{H^\pm}^2$. However, we are interested in the 
opposite limit, of sizeable $\tilde{m}/M$.}. 
As $\tilde{m}/M$ increases, the tree level Higgs
mass (or $\lambda$) also grows and this makes $\Delta_{\mu^2}$ decrease 
with $m_h$, just the opposite of the MSSM behaviour.
\begin{figure}
\vspace{1cm}
\centerline{
\vbox{
\psfig{figure=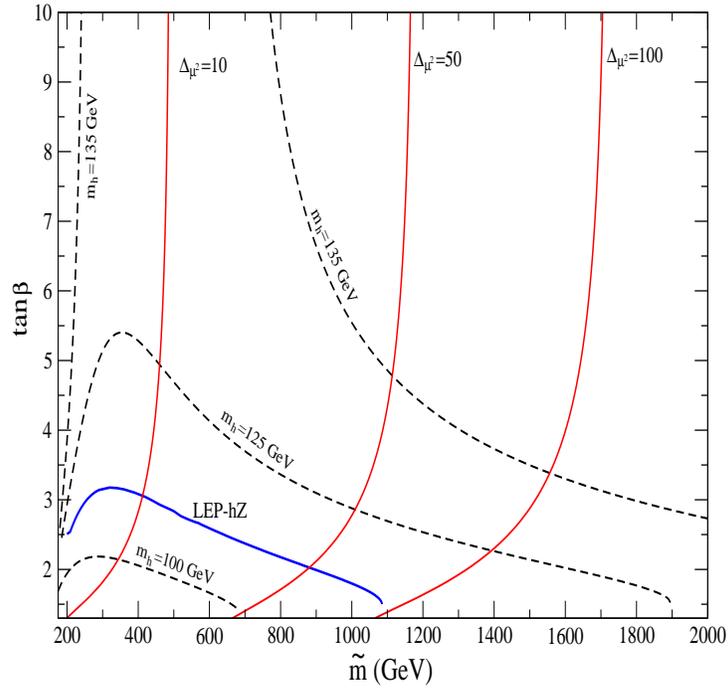,height=8cm,width=8cm,angle=-90,bbllx=6.cm,%
bblly=4.cm,bburx=21.cm,bbury=24.cm}
\vspace{1.75cm}
\psfig{figure=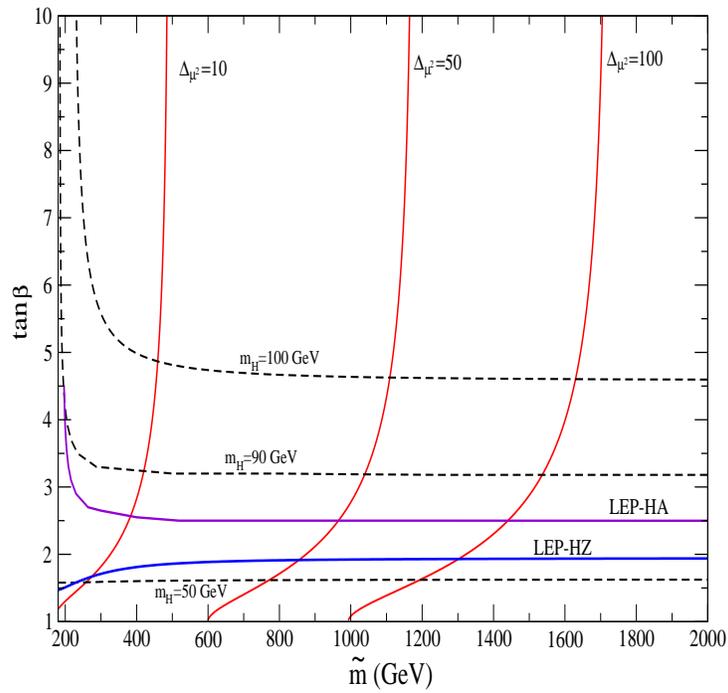,height=8cm,width=8cm,angle=-90,bbllx=4.cm,%
bblly=4.cm,bburx=19.cm,bbury=24.cm}
}}
\vspace{1.25cm}
\caption{\footnotesize
Fine tuning in the $(\tilde{m},\tan\beta)$ plane
in a low-scale SUSY breaking 
scenario with parameters as in set B. 
Dashed lines are contour lines of constant $m_h$ (upper plot) or $m_H$ 
(lower plot). The LEP bound for each case is also shown.}
\label{ftnew}
\end{figure}

Finally, fig.~\ref{ftnew} is the version of Fig.~\ref{ftMSSM} for this
unconventional model.  The values of the parameters are those of set B. We
show lines of constant $\Delta_{\mu^2}$ in the $(\tilde{m},\tan\beta)$
plane, together with lines of constant $m_h$ (upper plot) and $m_H$ (lower
plot).  In each plot we also draw the experimental lower bound on the
corresponding Higgs mass coming from LEP, either for Higgs-strahlung or 
associated production as indicated (see Appendix~B for details). We
find that the fine tuning is larger for smaller $\tan\beta$ and larger
$\tilde{m}$, as in the MSSM, but now the overall value of $\Delta_{\mu^2}$
is significantly smaller. From the figure we can estimate that for soft
masses $\tilde{m}^2\sim a v^2$, the fine tuning in this model (say near
$m_h=115$ GeV and $\tan\beta=3$) is $\Delta\sim 3.5 a$ instead of the 
$\sim 20a$ we found for the MSSM. The pattern of Higgs masses is also 
different
and restricting the fine tuning to be less than 10 does not
impose an upper bound on the Higgs masses, in contrast with the MSSM case.
As a result, the LEP bounds do not imply a large fine tuning: in the
region with small $\tilde{m}$ and $\tan\beta$ not too close to
1,\footnote{Besides the $\tan\beta>1$ region we have explored in this
paper, there is a wide region of parameter space with $\tan\beta=1$ which
is also experimentally allowed \cite{BCEN}.} we can get simultaneously 
Higgs masses
large enough to escape LEP detection and small fine tunings. In any case,
following the line of $\Delta_{\mu^2}=10$ we do find an upper bound
$\tilde{m}\simlt 500$ GeV, so that LHC would either find superpartners or
revive an (LHC) fine tuning problem for these scenarios (although the
problem would be much softer than in the MSSM).

\section{Concluding remarks}

\begin{enumerate}

\item As is well known, in the MSSM a successful electroweak
breaking requires substantial fine-tuning. This fine tuning is
abnormally large in the following sense: if the soft parameters have a
size $m_{\rm soft}^2\sim a v^2$, one expects a fine tuning of one
part in $a$; but in practice it is more than 20 times larger.

\item The main reason for that is the small
magnitude of the tree-level Higgs quartic coupling
$\lambda_{\rm MSSM}={1 \over 8}(g^2+g_Y^2)\cos^2 2\beta\ \simeq \  {1
\over 15}\cos^2 2\beta$. This has two effects:

\begin{itemize}

\item The ``natural'' value for the Higgs VEV, 
$v^2 \sim m_{\rm soft}^2/\lambda$ tends to be much larger 
than $m_{\rm soft}^2$, specially if $\tan\beta$ is not large.

\item Sizeable radiative corrections (and thus
sizeable soft terms) are needed to satisfy the experimental bound
on $m_h$, which worsens the fine tuning problem. Since $m_h$
increases logarithmically with $m_{\rm soft}$, the problem
gets exponentially worse for increasing $m_h$.

\end{itemize}

In addition, the radiative mechanism for EW breaking aggravates
the problem, since it induces  large coefficients for the individual
contributions of certain soft terms to the effective potential.

\item As a consequence, the most
efficient way of reducing the fine tuning is to consider
supersymmetric models where $\lambda_{\rm tree}$ is larger 
than in the MSSM. [An estimate of the expected improvement 
in the fine tuning, using the ordinary fine tuning parameters 
is given in eq.(\ref{Deltamu2s})]

\item The latter possibility takes place naturally in scenarios in
which the breaking of SUSY occurs at a low scale (not far from the TeV
scale). Then, the quartic couplings get \SUSY corrections,
$\delta \lambda\sim m_{\rm soft}^2/ M^2$,  so that  
$\lambda+\delta\lambda$ can be easily larger
than $\lambda_{\rm MSSM}$, as desired to ameliorate the fine tuning
problem. Moreover, this opens up many new 
possibilities for EW breaking and for a non-conventional Higgs 
spectrum.

\item We demonstrate this in an explicit model of low-scale \SUSY
studied in a previous paper by its own sake (and not with the goal of 
solving the fine tuning problem). This indicates that the improvement in 
the fine tuning is indeed a generic feature of these scenarios.

By modifying the parameters of the model we achieve a dramatic
improvement of the fine tuning for any range of $\tan \beta$ and
the Higgs mass (which can be as large as several hundred GeV if 
desired, but this is not necessary). It is in fact quite
easy to get e.g. $\Delta<5$ (i.e. no fine tuning), 
in contrast with the MSSM values, $\Delta>20$ (and much larger for 
$m_h>115$ GeV and/or small $\tan\beta$).

\item In scenarios with low-scale \SUSY, the interval of running of
the soft parameters is small, which has further consequences:

\begin{itemize}

\item EW breaking takes place at tree-level, which, as discussed in
point 2), also helps in reducing the fine tuning. 

\item The cross-talk (through RG running) between mass parameters
in the Higgs sector and those of other sectors (squarks, gluinos, etc.)
is drastically reduced. The latter can be (much) heavier than $M_Z$
without upsetting the naturalness of the electroweak scale. 
In this sense these scenarios represent an alternative to other options
which try to reduce the fine tuning by postulating correlations between 
different parameters to implement cancellations in $M_Z$: here $M_Z$
does not even depend strongly on those parameters.

\end{itemize}

\item The previous point is also very interesting for flavour physics in
two different fronts. First, in the MSSM the stringent FCNC bounds on the
non-universality of the sparticle mass matrices ({\em e.g.} from the
$K$-$\bar{K}$ system) could now be alleviated simply by increasing the
relevant soft masses ({\em e.g.} beyond 1 TeV) with negligible effect in
EW breaking.  Second, as it is known, even assuming UV universality, RG
evolution induces flavour violating effects which for the
$\mu\rightarrow e\gamma$ process are extremely dangerous. This problem is 
eliminated in the context of low scale SUSY breaking, where the RG effects 
are 
minimized (for a discussion see \cite{CIBarr}). Incidentally, the two 
flavour problems just mentioned can also be understood as fine tuning 
problems of the MSSM.

\item Finally, it is clear that, apart from scenarios with low-scale
\SUSY, there are other extensions of the MSSM which increase
$\lambda_{\rm tree}$ and thus improve the fine tuning. An
alternative, discussed in \cite{NMSSM}, is to enlarge the Higgs
sector, as in the NMSSM. This framework, 
however, is less
effective for a number of reasons. First, the simplest NMSSM model
gives an extra contribution to $\lambda$ that vanishes for large $\tan
\beta$, precisely the region  where the MSSM fine tuning was smallest
(even if still too large). Second, the conventional NMSSM with soft terms
generated at very high scale has important bounds on the previous extra
contribution, derived from the requirement of perturbativity. This
means that the available increase in $\lambda_{\rm tree}$ (and the
consequent improvement in the fine tuning) is more modest than could be
thought a priori. Finally, EW breaking still occurs radiatively, which
eliminates the extra bonus discussed in 6) above.

\end{enumerate}

\section*{A. General formulas for fine tuning parameters}
\setcounter{equation}{0}
\renewcommand{\theequation}{A.\arabic{equation}}

Here we consider a generic scenario where the Higgs sector consists 
of two $SU(2)_L$ doublets of opposite hypercharge, $H_1$ and $H_2$, as is
the case in many supersymmetric models \cite{HN}.
The most general Higgs potential 
for such two Higgs doublet models (2HDM) is (at tree level)
\bea
\label{V2HDMA}
V & = & m_1^2|H_1|^2  +
m_2^2|H_2|^2  - \left[ m_3^2 H_1\cdot H_2 + {\rm
h.c.}\right] \nonumber \\ & + &   {1\over 2}
\lambda_1|H_1|^4  +{1\over 2} \lambda_2|H_2|^4
+\lambda_3|H_1|^2 |H_2|^2  +\lambda_4|H_1\cdot
H_2|^2  \nonumber \\ & + &  \left[ {1 \over 2} \lambda_5
(H_1\cdot H_2)^2  +\lambda_6|H_1|^2 H_1\cdot H_2
+\lambda_7|H_2|^2 H_1\cdot H_2  +  {\rm h.c.}
\right]\ .
\eea
The minimum of this potential occurs in general at non-zero values of the 
neutral components of the Higgs doublets, $H_1^0$ and $H_2^0$ with
$\tan\beta\equiv \langle H^0_2 \rangle/\langle H^0_1 \rangle$ and 
$\langle H^0_1 \rangle=(v/\sqrt{2}) \cos\beta$, $\langle H^0_2 
\rangle=(v/\sqrt{2} )\sin\beta$. 
It is useful to write $V$ as a `SM-like'potential for $v$:
\be
\label{VbetaA}
V(v)={1 \over 2} m^2 v^2 + {1 \over 4} \lambda v^4 \ ,
\ee
where $\lambda$ and $m^2$ are functions of $\tan\beta$ and the initial 
parameters of the theory, $p_\alpha$. Explicitly
\be
\label{cimiA}
m^2 = \sum_{i=1}^3 c_i(\beta) m_i^2(p_\alpha), \;\; \;\;\; \;
\vec{c}=(c^2_\beta,s^2_\beta,-s_{2\beta})\ ,
\ee
and
\be
\label{dilambdaiA}
\lambda = \sum_{i=1}^7 d_i(\beta) \lambda_i(p_\alpha), \;\; \;\;\; \;
\vec{d}=({1\over 2}c^4_\beta,{1\over 2}s^4_\beta,s^2_\beta c^2_\beta,
s^2_\beta c^2_\beta,s^2_\beta c^2_\beta, 
c^2_\beta{s_{2\beta}}, s^2_\beta{s_{2\beta}})\ .
\ee
Minimization of $V$ with respect to $v$ and $\beta$ implies\footnote{With 
an abuse of notation we use the same symbols ($v$ and $\beta$) for the 
variables and their vacuum expectation values, but the meaning should be 
clear.} 
\be
\label{v2A}
v^2={-m^2\over \lambda} \ ,
\ee
\be
\label{tanbetaA}
2\lambda\frac{\partial m^2}{\partial \beta}
- m^2\frac{\partial \lambda}{\partial \beta}=0
\ .
\ee
In order to evaluate the fine tuning in a generic theory of this kind,
we will use the fine tuning parameters, $\Delta_{p_\alpha}$, introduced by 
Barbieri and Giudice \cite{BG}:  
\be
\label{BGA}
{\delta M_Z^2\over M_Z^2}= {\delta v^2\over v^2} = 
\Delta_{p_\alpha}{\delta
p_\alpha\over p_\alpha}\ ,  \ee
where $\delta M_Z^2$ ($\delta v^2$) is the change induced in $M_Z^2$
($v^2$) by a change $\delta p_\alpha$ in $p_\alpha$.
Naturalness requires $\Delta_{p_\alpha}\simlt {\cal O}(10)$.

Applying eq.~(\ref{BGA}) to eq.~(\ref{v2A}) we get, after trading 
$\partial m^2/\partial \beta$ by $\partial \lambda/\partial \beta$ 
using  eq.~(\ref{tanbetaA}),
\bea
\label{DeltapA}
\Delta_p = \frac{p}{m^2}\left[ \frac{\partial
m^2}{\partial p} + \frac{v^2}{2}\frac{\partial \lambda}{\partial
\beta}\frac{d \beta} {d p} + v^2\frac{\partial
\lambda}{\partial p} \right] \ .
\eea
The dependence of $\beta$ on $p$, which is not explicit
in the initial potential (\ref{V2HDMA}), can be extracted from
eq.~(\ref{tanbetaA}) by acting on it with $d/d p$, to obtain 
finally
\be
\label{Deltap2A}
\Delta_{p}=-{p\over x}\left[\left(2{\partial^{2} m^{2}\over \partial
\beta^{2}}+v^2{\partial^{2} \lambda\over \partial
\beta^{2}}\right)\left({\partial \lambda\over \partial p}+{1\over v^2}
{\partial m^{2}\over\partial p} \right)-{\partial \lambda\over \partial 
\beta}{\partial^{2} m^{2}\over \partial\beta\partial p} +{\partial 
m^{2}\over \partial \beta} {\partial^{2} \lambda\over
\partial\beta
\partial p} \right]\ ,
\ee
where
\begin{equation}
x\equiv 
\lambda\left(2\frac{\partial^{2}m^{2}}{\partial\beta^{2}}+
v^{2}\frac{\partial^{2}\lambda}{\partial\beta^{2}}\right)
-\frac{v^{2}}{2}\left(\frac{\partial\lambda}{\partial\beta}\right)^{2}\ .
\end{equation}
[Note that the dependence of $m^2$ and $\lambda$ on $\beta$
is determined by eqs.~(\ref{cimiA}, \ref{dilambdaiA}).]

In many cases, equations (\ref{DeltapA}) and (\ref{Deltap2A})
admit expansions which are useful for fine tuning estimates (although in 
the computations of this paper we have used the complete expressions).
If there exists a fine tuning at all, there must be some cancellation 
between
the various contributions to $m^2$, say ${\tt m_i^2}$, which generically 
implies
$\partial m^{2}/ \partial p= {\cal O}({\tt m_i^2}/ p)\gg 
{\cal O}(m^2/ p)$. Then, the last two terms
within the brackets in eq.~(\ref{DeltapA}) are suppressed
by a factor ${\cal O}(m^2/{\tt m_i^2})$, and
\bea
\label{DeltapaproxA}
\Delta_p \simeq \frac{p}{m^2}\frac{\partial
m^2}{\partial p} = -\frac{p}{\lambda v^2}\frac{\partial
m^2}{\partial p}\ .
\eea
The same result can be obtained from eq.~(\ref{Deltap2A}).

Let us now consider how the previous results are modified by radiative 
corrections. As is well-known, the 1-loop correction to the effective
Higgs potential in a supersymmetric theory (using the $\overline{\rm DR}$ 
renormalization scheme) is given by 
\bea
\label{V1loopA}
\delta_1V={1\over 64 \pi^2}\sum_a N_a M_a^4(H)\left[
\log{M_a^2(H)\over Q^2}-{3\over 2}\right]\ ,
\eea
where $Q$ is the renormalization scale, $M_a^2(H)$ is the $H$-dependent 
mass eigenvalue of the particle $a$ and $N_a$ its multiplicity 
(taken negative for fermions). 
$\delta_1V$ modifies the minimization conditions as well as $m_h$. 
However, it is possible
to reproduce these results by using appropriately (one-loop) corrected 
$m^2_i$,
$\lambda_i$ parameters in the tree-level expressions, e.g.
the minimization equations (\ref{v2A}, \ref{tanbetaA}) \cite{dCE}.
In this way, one can still use all
the previous (tree-level-like) equations (\ref{DeltapA}--\ref{DeltapaproxA})
for fine tuning estimates. In particular, the 
dominant contribution to the fine tuning is still given by 
eq.~(\ref{DeltapaproxA}) but expressed in terms of the one-loop corrected 
parameters.

Now, one expects $\delta_1 m_i^2
={\cal O}(N h^2 {\tilde m}^2/(32 \pi^2))$,
$\delta_1\lambda_i
={\cal O}(N h^4/(32 \pi^2))$, where $h$ is the coupling constant
of   a field with multiplicity $N$ to the Higgses
and ${\tilde m}^2$ is a typical soft mass. Moreover there can be 
a logarithmic factor $\sim \log({\tilde m}^2/m_t^2)$.
Clearly $\delta_1 m_i^2$ are smaller than the typical
${\cal O}({\tilde m}^2)$ tree-level contributions, so they do not
affect the degree of fine tuning. On the other hand, 
$\delta_1\lambda_i$ can be relevant if the tree-level
values are small, as it happens for instance in the MSSM
(but not in models with sizeable $\lambda_{\rm tree}$,
as those considered in this paper).
These corrections are normally dominated by the top-stop sector with
coupling $h_t=\sqrt{2}m_t/(v \sin\beta)$, which besides being 
${\cal O}(1)$ has large
multiplicity, $N_L+N_R=12$. If some of the Higgs self-couplings, 
$\lambda_i$,
are initially large, say ${\cal O}(1)$, they can also contribute
substantially to $\delta_1\lambda_i$, though the multiplicity is 
smaller than for the stops. However, as mentioned above, in this case 
$\delta_1\lambda_i\ll \lambda_{\rm tree}$ and therefore such corrections 
can be ignored for the fine tuning issue.

Consequently, for fine tuning estimates, we approximate the radiative 
corrections by the logarithmic stop contribution
(more sophisticated expressions for $\delta_1\lambda_i$ can be found in 
\cite{CEQWHHH}):
\be
\label{lambdaradA}
\delta_1\lambda_2={3h_t^4\over 8\pi^2}\log{M_{\rm SUSY}^2\over m_t^2}\ .
\ee
In particular the approximate formula given in eq.~(\ref{DeltapaproxA})
simply gets corrected by a factor 
$\lambda_{\rm tree}/\lambda_{\rm 1-loop}$.

\section*{B. LEP Higgs bounds}
\setcounter{equation}{0}
\renewcommand{\theequation}{B.\arabic{equation}}

The main Higgs production mechanism
of the physical  CP-even scalars ${\cal H}_\alpha^0=h^0, H^0$ at LEP is 
$e^+e^-\rightarrow Z^0{\cal H}_\alpha^0$. 
The Higgs production cross-section is
\be
\sigma_{Z{\cal 
H}_\alpha}=\xi_{{\cal H}_\alpha}^2 \sigma_{Zh}^{SM}(m^2_{{\cal 
H}_\alpha})\ , \ee
where $\sigma_{Zh}^{SM}(m^2)$ is the SM production cross-section
for a Higgs with mass $m$ \cite{SZ} and the prefactor $\xi_{{\cal 
H}_\alpha}$ 
measures
the coupling $Z^0Z^0{\cal H}_\alpha^0$ relative to the SM value. In a 
generic 2HDM the linear combination  along the breaking direction\footnote{We 
write $H_1^0=(v_1+h_1^{0r}+ih_1^{0i})/\sqrt{2}$ and a similar formula for 
$H_2^0$.}
$h_\parallel\equiv h_1^{0r}\cos\beta+h_2^{0r}\sin\beta$
has a coupling to $ZZ$ of
SM strength while the orthogonal combination $h_\perp\equiv
h_1^{0r}\sin\beta-h_2^{0r}\cos\beta$
does not couple to $ZZ$. In the basis $\{h_\parallel,h_\perp\}$
the mass eigenstates $h^0,H^0$ read
\be
h^0=\xi_h h_\parallel + \xi_H h_\perp\ ,\;\;\;\;
H^0=\xi_H h_\parallel - \xi_h h_\perp\ ,
\ee
with $\xi_h^2+\xi_H^2=1$. That is, the coupling ${\cal H}_\alpha^0Z^0Z^0$
is proportional to the amount of $h_\parallel$ that enters in the
composition of ${\cal H}_\alpha^0$. From the definition of
$\tan\beta$ and that of 
the mixing angle of the two CP-even Higgs bosons $h^0,H^0$:
\bea
h^0&=&h_2^{0r}\cos\alpha-h_1^{0r}\sin\alpha\ ,\nonumber\\
H^0&=&h_1^{0r}\cos\alpha+h_2^{0r}\sin\alpha\ ,
\eea
we obtain the familiar expressions
\be
\xi_h^2=\sin^2(\alpha-\beta)\ ,
\;\;
\;\;
\xi_H^2=\cos^2(\alpha-\beta)\ .
\ee

In the alternative scenario considered in section~4, at tree level, the 
mass matrix for CP-even Higgses (in the basis $\{h_1^{0r},h_2^{0r}\}$) is
\bea
M^2_{{\cal H}_\alpha^0}&=&\left[
\begin{array}{ccc}
m_\parallel^2c^2_\beta+m_\perp^2s^2_\beta & &
(m_\parallel^2-m_\perp^2)c_\beta s_\beta \\
& & \\
(m_\parallel^2-m_\perp^2)c_\beta s_\beta & &
m_\parallel^2 s^2_\beta+m_\perp^2 c^2_\beta
\end{array}
\right]\nonumber\\
&&\nonumber\\
&=&\left(
\begin{array}{cc}
c_\beta & s_\beta \\
& \\
s_\beta & -c_\beta
\end{array}
\right)\left[
\begin{array}{cc}
m_\parallel^2 & 0 \\
& \\
0 &
m_\perp^2 \end{array}
\right]
\left(
\begin{array}{cc}
c_\beta & s_\beta \\
& \\
s_\beta & -c_\beta
\end{array}
\right)
\ .
\label{massmatrix}
\eea
This implies that $h_\parallel$ and $h_\perp$ are in fact mass 
eigenstates
and means in particular that
only $h_\parallel$ could be produced at LEP. For some choice of 
parameters (like in set A, used in section 4),
$h_\parallel$ turns out to be the heavy state, and its mass makes it
kinematically inaccessible at LEP. The light state turns out to be
$h_\perp$ and even if it is light it does not couple to $Z^0$ and
therefore it is not produced.
                                                                                
At one loop the previous situation changes. Corrections to the mass matrix
(\ref{massmatrix}), the main one being
\be
\delta\langle h_2^{0r}|M^2_{H_i^0}|h_2^{0r}\rangle={3 m_t^4\over
\pi^2v^2s_\beta^2}\log{M_{\rm SUSY}\over m_t}\ ,
\ee
induce deviations of the mass eigenstates from $h_\parallel$ and
$h_\perp$, making $\xi_h$ and $\xi_H$ different from 1 and 0.
Working out the expression for the one-loop corrected $\alpha$ we arrive
at the simple result
\be
\xi_h^2={\displaystyle{(m^2_H-m^2_{h_\parallel})c_\beta^2
+(m^2_{h_\perp}-m^2_h)s_\beta^2
\over m^2_H-m^2_h}}\ ,
\ee
and
\be
\xi_H^2={\displaystyle{(m^2_H-m^2_{h_\perp})s_\beta^2
+(m^2_{h_\parallel}-m^2_h)c_\beta^2
\over m^2_H-m^2_h}}\ .
\ee
As discussed before, $\{h_\parallel,h_\perp\}$ are the tree level
mass eigenstates.

In order to implement the LEP bound in this alternative scenario, we 
conservatively impose that
$\sigma_{Z{\cal H}_\alpha}$ should be smaller than 
$\sigma_{Zh}^{SM}(m^2_{h})$
evaluated at $\sqrt{s}=209$ GeV and $m_H=115$ GeV (the ultimate LEP bound
on the SM Higgs mass).  This requirement can be
represented as an upper limit on $\xi_{{\cal H}_\alpha}^2$ as a function 
of $m_h$.
A more refined bound (unnecessarily sophisticated for our purposes)
can be found on the experimental papers \cite{LEPa,LEPb}.

Another possible Higgs production mechanism is associated production 
$e^+e^-\rightarrow A^0{\cal H}_\alpha^0$, with cross section given by
\cite{SZ}
\be
\sigma_{A{\cal H}_\alpha}=(1-\xi_{{\cal H}_\alpha}^2)\bar\lambda
\sigma_{Zh}^{SM}(m^2_{{\cal 
H}_\alpha})\ , 
\ee
where $\bar\lambda$ is a kinematical factor. The non observation of this 
process sets a limit on our model. We implement this limit by using the 
experimental bound on the coefficient $(1-\xi_{{\cal H}_\alpha}^2)$ 
derived {\em e.g} in \cite{Ah} as a function of $m_{{\cal 
H}_\alpha}+ m_A$. (We are conservative in 
using that experimental curve, which applies strictly to the 
case $m_{{\cal H}_\alpha}\simeq m_A$, and in assuming $\sim 
100\%$ branching ratios $A\rightarrow b\bar b$ and 
${\cal H}_\alpha\rightarrow b\bar 
b$.) When $(1-\xi_{{\cal H}_\alpha}^2)\simeq 1$, this 
limit reads $m_{{\cal
H}_\alpha}+ m_A\simlt 195$ GeV.

Finally, charged Higgs production ($e^+e^-\rightarrow H^+H^-$) does not 
give  constraints in this scenario because $m_{H^\pm}\simeq 95$ GeV while 
the experimental limit is around $m_{H^\pm}\simlt 80$ GeV \cite{Hpm}.

\section*{Acknowledgments}
We thank Andrea Brignole and Ignacio Navarro for stimulating discussions 
and an anonymous referee for a most useful report which led to an 
improved final version of the paper.  
This work is supported in part by the Spanish Ministry of Science and 
Technology through a MCYT project (FPA2001-1806).
The work of I.~Hidalgo has been supported by a FPU grant from the Spanish
Ministry of Education Culture and Sport. J.A.~Casas thanks the IPPP
(Durham) for the hospitality during the final stages of this work and 
J.R.~Espinosa thanks the CERN TH Division for hospitality during the 
initial stages of this work.



\begin{thebibliography}{99}
%
\bibitem{SUSY}
H.~P.~Nilles,
Phys.\ Rept.\  {\bf 110} (1984) 1;
H.~E.~Haber and G.~L.~Kane,
Phys.\ Rept.\  {\bf 117} (1985) 75;
S.~Weinberg,
``The Quantum Theory Of Fields.  Vol. 3: Supersymmetry'',
Cambridge, UK: Univ. Pr. (2000).
%
\bibitem{Ibanez}
L.~E.~Ib\'a\~nez and G.~G.~Ross,
Phys.\ Lett.\ B {\bf 110} (1982) 215;
K.~Inoue, A.~Kakuto, H.~Komatsu and S.~Takeshita,
Prog.\ Theor.\ Phys.\  {\bf 68} (1982) 927
[Erratum-ibid.\  {\bf 70} (1983) 330];
L.~E.~Ib\'a\~nez and C.~L\'opez,
Nucl.\ Phys.\ B {\bf 233} (1984) 511;
L.~E.~Ib\'a\~nez, C.~L\'opez and C.~Mu\~noz,
Nucl.\ Phys.\ B {\bf 256} (1985) 218.
%
\bibitem{BG}
R.~Barbieri and G.~F.~Giudice,
Nucl.\ Phys.\ B {\bf 306} (1988) 63.
%
\bibitem{dCC}
B.~de Carlos and J.~A.~Casas,
Phys.\ Lett.\ B {\bf 309} (1993) 320
[hep-ph/9303291].
%
\bibitem{Poko}
M.~Olechowski and S.~Pokorski,
Nucl.\ Phys.\ B {\bf 404} (1993) 590
[hep-ph/9303274].
%
\bibitem{Anderson}
G.~W.~Anderson and D.~J.~Casta\~no,
Phys.\ Lett.\ B {\bf 347} (1995) 300
[hep-ph/9409419];
Phys.\ Rev.\ D {\bf 52} (1995) 1693
[hep-ph/9412322];
Phys.\ Rev.\ D {\bf 53} (1996) 2403
[hep-ph/9509212].
%
\bibitem{toni}
G.~W.~Anderson, D.~J.~Castano and A.~Riotto,
Phys.\ Rev.\ D {\bf 55} (1997) 2950 [hep-ph/9609463].
%
\bibitem{Paolo}
P.~Ciafaloni and A.~Strumia,
Nucl.\ Phys.\ B {\bf 494} (1997) 41
[hep-ph/9611204].
%
\bibitem{LEPprob}
P.~H.~Chankowski, J.~R.~Ellis and S.~Pokorski,
Phys.\ Lett.\ B {\bf 423} (1998) 327
[hep-ph/9712234].
%
\bibitem{LEPprob2}
R.~Barbieri and A.~Strumia,
Phys.\ Lett.\ B {\bf 433} (1998) 63
[hep-ph/9801353].
\bibitem{CEOP}
P.~H.~Chankowski, J.~R.~Ellis, M.~Olechowski and S.~Pokorski,
Nucl.\ Phys.\ B {\bf 544} (1999) 39
[hep-ph/9808275].
%
\bibitem{Kane}
G.~L.~Kane and S.~F.~King,
Phys.\ Lett.\ B {\bf 451} (1999) 113
[hep-ph/9810374].
%
\bibitem{mar}
M.~Bastero-Gil, G.~L.~Kane and S.~F.~King,
Phys.\ Lett.\ B {\bf 474} (2000) 103
[hep-ph/9910506].
%
\bibitem{rest}
S.~Dimopoulos and G.~F.~Giudice,
Phys.\ Lett.\ B {\bf 357} (1995) 573
[hep-ph/9507282];
K.~Agashe and M.~Graesser,
Nucl.\ Phys.\ B {\bf 507} (1997) 3
[hep-ph/9704206];
%
L.~Giusti, A.~Romanino and A.~Strumia,
Nucl.\ Phys.\ B {\bf 550} (1999) 3
[hep-ph/9811386];
%
J.~L.~Feng, K.~T.~Matchev and T.~Moroi,
Phys.\ Rev.\ Lett.\  {\bf 84} (2000) 2322
[hep-ph/9908309];
%
K.~Agashe,
Phys.\ Rev.\ D {\bf 61} (2000) 115006
[hep-ph/9910497];
%
A.~Romanino and A.~Strumia,
Phys.\ Lett.\ B {\bf 487} (2000) 165
[hep-ph/9912301].
%
\bibitem{Howie} See {\em e.g.} the appendix of
J.~A.~Casas, J.~R.~Espinosa and H.~E.~Haber,
Nucl.\ Phys.\ B {\bf 526} (1998) 3
[hep-ph/9801365].
%
\bibitem{Gordy}
G.~L.~Kane, J.~Lykken, B.~D.~Nelson and L.~T.~Wang,
Phys.\ Lett.\ B {\bf 551} (2003) 146
[hep-ph/0207168].
%
\bibitem{LEPbound}
[ALEPH, DELPHI, L3 and OPAL Collaborations],
Phys.\ Lett.\ B {\bf 565} (2003) 61
[hep-ex/0306033].
%
\bibitem{mhrad}
J.~Ellis, G.~Ridolfi and F.~Zwirner,
Phys.\ Lett.\ B {\bf 257} (1991) 83;
Phys.\ Lett.\ B {\bf 262} (1991) 477;
Y.~Okada, M.~Yamaguchi, and T.~Yanagida,
Prog.\ Theor.\ Phys.\ {\bf 85} (1991) 1;
Phys.\ Lett.\ B {\bf 262} (1991) 54;
H.~E.~Haber and R.~Hempfling,
Phys.\ Rev.\ Lett.\ {\bf 66} (1991) 1815;
R.~Barbieri, M.~Frigeni and M.~Caravaglios,
Phys.\ Lett.\ B {\bf 258} (1991) 167.
%
\bibitem{BCEN}
A.~Brignole, J.~A.~Casas, J.~R.~Espinosa and I.~Navarro,
[hep-ph/0301121].
%
\bibitem{Gordy2}
G.~L.~Kane, B.~D.~Nelson, T.~T.~Wang and L.~T.~Wang,
[hep-ph/0304134].
%
\bibitem{top}
F.~Abe {\it et al.}  [CDF Collaboration],
Phys.\ Rev.\ Lett.\  {\bf 82} (1999) 271
[Erratum-ibid.\  {\bf 82} (1999) 2808]
[hep-ex/9810029];
B.~Abbott {\it et al.}  [D0 Collaboration],
Phys.\ Rev.\ D {\bf 60} (1999) 052001
[hep-ex/9808029].
%
\bibitem{hard}
K.~Harada and N.~Sakai,
Prog.\ Theor.\ Phys.\  {\bf 67} (1982) 1877;
D.~R.~Jones, L.~Mezincescu and Y.~P.~Yao,
Phys.\ Lett.\ B {\bf 148} (1984) 317;
I.~Jack and D.~R.~Jones,
Phys.\ Lett.\ B {\bf 457} (1999) 101
[hep-ph/9903365];
L.~J.~Hall and L.~Randall,
Phys.\ Rev.\ Lett.\  {\bf 65} (1990) 2939;
F.~Borzumati, G.~R.~Farrar, N.~Polonsky and S.~Thomas,
Nucl.\ Phys.\ B {\bf 555} (1999) 53
[hep-ph/9902443];
S.~P.~Martin,
Phys.\ Rev.\ D {\bf 61} (2000) 035004
[hep-ph/9907550].
%
\bibitem{Brignole}
A.~Brignole, F.~Feruglio and F.~Zwirner,
Nucl.\ Phys.\ B {\bf 501} (1997) 332
[hep-ph/9703286].
%
\bibitem{Polonsky}
N.~Polonsky and S.~Su,
Phys.\ Lett.\ B {\bf 508} (2001) 103
[hep-ph/0010113];
Phys.\ Rev.\ D {\bf 63} (2001) 035007
[hep-ph/0006174].
%
\bibitem{Strumia}
A.~Strumia,
Phys.\ Lett.\ B {\bf 466} (1999) 107 [hep-ph/9906266].
%
\bibitem{extG} 
See {\em e.g.} 
D.~Comelli and C.~Verzegnassi,
Phys.\ Lett.\ B {\bf 303} (1993) 277;
J.~R.~Espinosa and M.~Quir\'os,
Phys.\ Lett.\ B {\bf 302} (1993) 51
[hep-ph/9212305];
M.~Cveti\v c, D.~A.~Demir, J.~R.~Espinosa, L.~L.~Everett and P.~Langacker,
Phys.\ Rev.\ D {\bf 56} (1997) 2861
[Erratum-ibid.\ D {\bf 58} (1998) 119905]
[hep-ph/9703317].
P.~Batra, A.~Delgado, D.~E.~Kaplan and T.~M.~Tait,
[hep-ph/0309149].
%
\bibitem{extH}
M.~Drees,
Int.\ J.\ Mod.\ Phys.\ A {\bf 4} (1989) 3635;
J.~R.~Ellis, J.~F.~Gunion, H.~E.~Haber, L.~Roszkowski and F.~Zwirner,
Phys.\ Rev.\ D {\bf 39} (1989) 844;
P.~Binetruy and C.~A.~Savoy,
Phys.\ Lett.\ B {\bf 277} (1992) 453.
J.~R.~Espinosa and M.~Quir\'os,
Phys.\ Lett.\ B {\bf 279} (1992) 92;
Phys.\ Rev.\ Lett.\  {\bf 81} (1998) 516
[hep-ph/9804235];
G.~L.~Kane, C.~F.~Kolda and J.~D.~Wells,
Phys.\ Rev.\ Lett.\  {\bf 70} (1993) 2686
[hep-ph/9210242].
%
\bibitem{NMSSM}
M.~Bastero-Gil, C.~Hugonie, S.~F.~King, D.~P.~Roy and S.~Vempati,
Phys.\ Lett.\ B {\bf 489} (2000) 359
[hep-ph/0006198].
%
\bibitem{CIBarr}
J.~A.~Casas and A.~Ibarra,
Nucl.\ Phys.\ B {\bf 618} (2001) 171
[hep-ph/0103065];
S.~M.~Barr,
[hep-ph/0307372].
%
\bibitem{HN}
H.~E.~Haber and Y.~Nir,
Nucl.\ Phys.\ B {\bf 335} (1990) 363.
%
\bibitem{CEQWHHH}
M.~Carena, J.~R.~Espinosa, M.~Quir\'os and C.~E.~Wagner,
Phys.\ Lett.\ B {\bf 355} (1995) 209
[hep-ph/9504316];
H.~E.~Haber, R.~Hempfling and A.~H.~Hoang,
Z.\ Phys.\ C {\bf 75} (1997) 539
[hep-ph/9609331].
%
\bibitem{dCE}
See {\em e.g.} sect.~3 in
B.~de Carlos and J.~R.~Espinosa,
Nucl.\ Phys.\ B {\bf 503} (1997) 24 [hep-ph/9703212].
%
\bibitem{SZ}
See {\em e.g.} M.~Spira and P.~M.~Zerwas,
[hep-ph/9803257].
%
\bibitem{LEPa}
A.~Heister {\it et al.}  [ALEPH Collaboration],
Phys.\ Lett.\ B {\bf 526} (2002) 191
[hep-ex/0201014];
J.~Fernandez  [DELPHI Collaboration],
[hep-ex/0307002];
G.~Abbiendi {\it et al.}  [OPAL Collaboration],
Eur.\ Phys.\ J.\ C {\bf 27} (2003) 311
[hep-ex/0206022];
P.~Achard {\it et al.}  [L3 Collaboration],
Phys.\ Lett.\ B {\bf 545} (2002) 30
[hep-ex/0208042].
%
\bibitem{LEPb}
G.~Abbiendi {\it et al.}  [OPAL Collaboration],
Eur.\ Phys.\ J.\ C {\bf 18} (2001) 425
[hep-ex/0007040].
%
\bibitem{Ah}
[LEP Higgs Working Group Collaboration],
[hep-ex/0107030].
%
\bibitem{Hpm}
[LEP Higgs Working Group for Higgs boson searches Collaboration],
[hep-ex/0107031].
%
\end{thebibliography}
\end{document}